\documentclass[10pt,letterpaper,compsoc,conference]{template/iiswc26}

\usepackage{cite}
\usepackage{amsmath,amssymb,amsfonts}
\usepackage{algorithmic}
\usepackage{graphicx}
\usepackage[dvipsnames,table]{xcolor}
\usepackage[final]{microtype}
\usepackage[italic]{mathastext}
\usepackage{libertine}
\usepackage[T1]{fontenc}
\usepackage{textcomp}
\usepackage[varqu,varl]{zi4}
\usepackage[all]{template/nowidow}
\usepackage[keeplastbox]{template/flushend}
\usepackage{fancyhdr}
\usepackage{harveyballs}

\usepackage{booktabs}
\usepackage{listings}
\usepackage{subcaption}
\usepackage{svg}
\usepackage{float}
\usepackage{mdframed}
\usepackage{xspace}
\usepackage{url}
\usepackage{makecell}
\usepackage[most]{tcolorbox}
\usepackage{tikz}
\usepackage{placeins}
\usepackage{tabulary}
\usepackage{wrapfig}
\usepackage[hidelinks]{hyperref}

\definecolor{verylightgray}{gray}{0.9}

\lstdefinestyle{compactPython}{
  language=Python,
  basicstyle=\footnotesize\ttfamily,
  keywordstyle=\color{blue},
  commentstyle=\color{gray},
  frame=single,
  breaklines=true,
  captionpos=b,
  aboveskip=0.3em,
  belowskip=0.3em,
  lineskip=0pt,
  xleftmargin=1.5em,
  xrightmargin=0em,
  framexleftmargin=1.5em,
  framexrightmargin=0em,
  numbers=left,
  numberstyle=\tiny\color{gray},
  numbersep=1em,
  showstringspaces=false,
  linewidth=\columnwidth
}

\newfloat{cascade}{hbtp}{lop}
\floatname{cascade}{Cascade}

\definecolor{pastelpink}{RGB}{250,190,233}
\definecolor{pastelyellow}{RGB}{250,246,186}
\definecolor{pastelpurple}{RGB}{221,199,255}
\definecolor{pastelblue}{RGB}{174,198,207}

\AtBeginDocument{%
  }

\tcbset{textmarker/.style={%
        enhanced,
        parbox=false,boxrule=0mm,boxsep=0mm,arc=0mm,
        outer arc=0mm,left=4mm,right=3mm,top=7pt,bottom=7pt,
        toptitle=1mm,bottomtitle=1mm,oversize}}

\newtcolorbox[auto counter]{sidebar_box}[2][]{textmarker,
    breakable,
    borderline west={6pt}{0pt}{pastelblue},
    colback=pastelblue!10!white,
    halign=justify,
    title=\textcolor{black}{\textbf{Inset~\thetcbcounter}~#2},
    title code={
      \path[fill=pastelblue!10!white] (title.south west) rectangle (title.north east);
      \path[draw=pastelblue,solid,line width=0.75mm]
      ([xshift=0mm]title.south west) -- ([xshift=0mm]title.south east);
      },
    nameref={#2},
    #1
}

\begin{document}

\title{Campaign Diagrams: Visualizing the March Through the Phases of a Workload}

\author{%
  \IEEEauthorblockN{Toluwanimi O. Odemuyiwa\IEEEauthorrefmark{1},
                    John D. Owens\IEEEauthorrefmark{1},
                    Michael Pellauer\IEEEauthorrefmark{2},
                    Joel S. Emer\IEEEauthorrefmark{2}\IEEEauthorrefmark{3}}
  \IEEEauthorblockA{\IEEEauthorrefmark{1}University of California, Davis\quad
                    \IEEEauthorrefmark{2}NVIDIA\quad
                    \IEEEauthorrefmark{3}Massachusetts Institute of Technology}
}

\maketitle
\fancypagestyle{numbered}{%
  \fancyhf{}%
  \renewcommand{\headrulewidth}{0pt}%
  \fancyfoot[C]{\thepage}%
}
\thispagestyle{numbered}
\pagestyle{numbered}

\begin{abstract}
We present campaign diagrams, a visualization technique for phase-level analysis of resource utilization and bottlenecks in modern workloads.
Existing tools have a trade-off: rooflines aggregate a workload into a single point and lose all notion of time, while profilers and traces expose fine-grained events but obscure what bounds performance. Instead, a campaign diagram depicts compute throughput and memory bandwidth utilization, compute and memory traffic volume, and latency in a single figure. Since they can be generated from analytical models, simulations, or profiling data, campaign diagrams capture both ideal bounds and a kernel's achieved performance. We demonstrate them on two case studies: a low-rank GEMM, where they reveal the counterintuitive result that reducing operational intensity can improve end-to-end performance, and Mamba, where they expose fusion and pipelining opportunities across phases. In both cases, our visualization technique reveals optimization opportunities that are difficult to identify with rooflines or profilers alone.

\end{abstract}

\section{Introduction}
Implementing modern workloads requires exploring a rich design space, including algorithm selection, data format, data and work partitioning, fusion, and dataflow, among others~\cite{Ragan-Kelley:2013:HAL, Kjolstad:2017:TAC, Nayak:2023:TDF_micro, odemuyiwa:2026:edge}.
In this space, different choices lead to potentially drastic differences in performance and overall resource efficiency.
To evaluate the impact of a design decision, researchers need answers to the following simple, but critical, questions:
\begin{enumerate}
\item Which parts of the workload are resource-bound and which resource (e.g., memory? compute?) bounds each part?
\item Throughout the execution lifetime of the workload, how does resource utilization evolve?
\item Where are the latency bottlenecks? Which phases are worth optimizing (and which are already saturated)?
\item How do different implementations compare in latency, resource utilization, and compute and memory volume? What causes these differences?
\end{enumerate}

\begin{figure}
    \centering
\includegraphics[width=\linewidth]{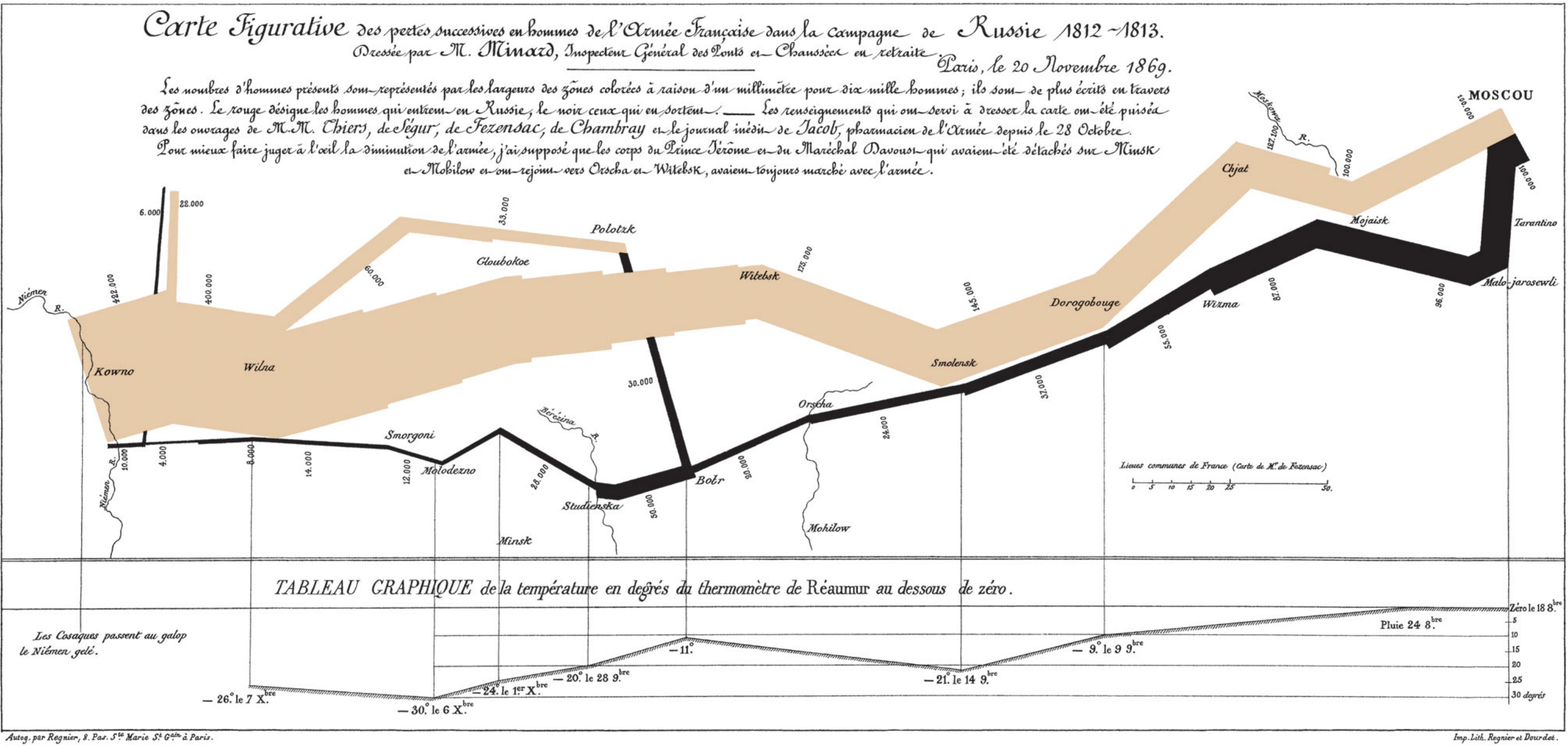}
  \vspace{-0.6em}
  \caption{Napoleon's March as Visualized by Charles Minard~\cite{tufte:1983:VDQ}, the inspiration behind our proposed campaign diagrams.}\label{fig:march}
\end{figure}

The roofline model~\cite{williams:2009:RIV, hill:2019:gables} is a seminal tool for addressing these questions.
It plots compute throughput (y-axis) against \emph{operational intensity} (number of compute operations per byte of data movement, x-axis).
A ``roofline'' represents architectural limits on memory bandwidth and compute throughput, separating the plot into memory bandwidth-bound regions and compute throughput-bound regions.
This provides insight into the hardware limitations or implementation shortcomings of a given workload.
Fundamentally, however, roofline plots provide an \emph{aggregate} view of either portions of the workload or the entire workload.
Execution over a given timescale is compressed into a single point that represents operational intensity (OI), memory bandwidth utilization, and compute throughput.
Choosing an appropriate timescale is difficult: if too fine-grained, the roofline becomes cluttered as many points overlap; and if too coarse (e.g., over the entire execution of the workload), performance insights collapse (see \S~\ref{sssec:rooflines}).
This obscures performance characteristics over the lifetime of the workload.
Modern, valuable workloads such as Deep Neural Networks or Large Language Models are not monolithic: they consist of distinct phases of computation, each with specific data access patterns, compute, and resource demands.
As such, optimizing a given workload requires reasoning not only about the overall workload performance, but also about the \emph{individual} phases comprising it and its behavior over time.
Rooflines cannot provide insight into (1) the duration of each phase in a workload, (2) when bottlenecks occur, and (3) how implementations with similar operational intensities or resource utilization differ in behavior over time.
Meanwhile, popular trace-based visualizations may show fine-grained operations and when they occur, but do not provide easy insight into \emph{what} is bounding performance.
An effective visualization should expose (1) the compute and memory bandwidth resource utilization and volume at (2) the component level of a workload (3) over time.
Without this perspective, it remains difficult to identify which phases dominate execution time, how resource utilization evolves throughout execution, and where to focus optimization efforts.

To address these shortcomings, we present \emph{campaign diagrams}.
We are inspired by Charles Joseph Minard's visualization of Napoleon's march in the Russian campaign~\cite{tufte:1983:VDQ} (Figure~\ref{fig:march}).
Minard's visualization combines multiple dimensions (time, distance traveled, direction of travel, troop size) into a single coherent narrative of the campaign's failure as time progressed.
Similarly, campaign diagrams visualize the ``march'' of a workload through its phases.
Rather than collapsing execution into a single aggregate point, a campaign diagram combines memory and compute volume, memory bandwidth utilization, compute throughput, and execution time into a single visualization.

\paragraph{Contributions}
Overall, we make the following contributions:
\begin{itemize}
    \item We present \emph{campaign diagrams}, a visualization technique that combines workload phases, compute and memory volume, compute utilization, memory-bandwidth utilization, and execution time into a single view (\S~\ref{sec:simple-campaign}).\footnote{We provide an implementation in Altair~\cite{vanderplas:2018:altair} that allows users to interact dynamically with the plots.}
    \item Through case studies, we show how campaign diagrams reveal (1) optimization opportunities for concurrent execution of phases (pipelining), (2) the impact of fusion on the \emph{volume} of memory traffic and amount of computational work (\S~\ref{ssec:simple-fusion}), and (3) dilation and throttling effects when pipelining oversubscribes resources (\S~\ref{sec:pipeline}).
    \item We show that campaign diagrams reveal counterintuitive algorithmic choices. For example, \emph{reducing} operational intensity can \emph{improve} overall performance, because campaign diagrams expose compute and memory volume directly (\S~\ref{sec:case-lr-gemm}).
\end{itemize}

\section{Visualization Goals}\label{sec:goals}
The following high-level goals guide our approach:

\begin{enumerate}
  \item\label{goal:agnostic} \textbf{Execution-trace agnostic.} We should be able to visualize the results of real runs, analytical models, and everything in between.
  The system should produce diagrams from phase-level metrics, regardless of the original source.

  \item\label{goal:phase-level} \textbf{Reveal phase-level metrics.} Resource utilization and volume across phases should be clear. A user should be able to tell which resources are underutilized for each phase, which resources are causing bottlenecks, and which phases are bottlenecked. In this paper, we focus on compute throughput and memory bandwidth, but similar visual encodings apply to any pair of resources with measurable utilization and volume.

  \item\label{goal:suggest-opts} \textbf{Suggest optimization opportunities.} Given the availability of resources at any given point in time, and the user's domain-specific knowledge (e.g., phase execution dependencies, etc.), the diagrams should reveal potential regions of time where fusion, dilation and throttling, or concurrent execution may help.

  \item\label{goal:time-scale} \textbf{Time-scale agnostic.} The visualization and analysis should be agnostic to the time scale (e.g., cycles, seconds, or abstract ``ticks''). The granularity of a ``phase'' is determined upstream.\footnote{In this paper we use a sequence of tensor algebra computations (i.e., \emph{Einsums}~\cite{Einstein:1916:FGT,Kjolstad:2017:TAC,odemuyiwa:2026:edge}), roughly comparable to a single layer of a neural network as the granularity for most of the examples.}

  \item\label{goal:platform-agnostic} \textbf{Platform-agnostic.} The diagrams should not be tied to any particular underlying architecture, hence we use fractions of peak resource utilization. Thus, the visual encoding is applicable across architectures as long as the relevant peak resource rates are known.
\end{enumerate}

Additionally, we have the following non-goals:
\begin{enumerate}
  \item\label{nongoal:profilers} We are not trying to replace fine-grained profilers; these tools do not provide details on instruction counts, miss rates, or hardware performance counters.
  \item\label{nongoal:prescribe} We are not prescribing optimization techniques; the optimizations revealed in the case studies are patterns that can help a campaign diagram user notice opportunities.
  \item\label{nongoal:num_resources} We do not focus on plotting more than two resource types simultaneously~\cite{checconi:2022:ridgeline}. In this paper, we focus on memory and compute resources for comparison with standard roofline plots. However, \textbf{our technique works with other pairs of rate-limited resources}.
\end{enumerate}

Note that all of the above restrictions are the same as those imposed by roofline diagrams, which have proven to be a highly successful level of abstraction. This is a purposeful decision: readers familiar with roofline visualizations should already be comfortable with them, and we impose no further conditions.

\section{Existing Approaches}\label{ssec:limits}
Table~\ref{tab:related-works} motivates our contribution compared to various approaches to performance visualization. We discuss them in detail below.

\begin{table*}
\centering
\caption{Campaign diagrams complement existing approaches by supporting phase-aware reasoning at design time without sacrificing abstraction.}
\label{tab:related-works}
\setlength{\tabcolsep}{2pt}
\renewcommand{\arraystretch}{1.0}
\begin{scriptsize}
\begin{tabulary}{\textwidth}{LLLLL}
\toprule
\textbf{Attribute} &
\textbf{Roofline} &
\textbf{Hierarchical Roofline} &
\textbf{Runtime Profilers~\cite{MediaTek:2024:NeuroPilotSDK, Arm:2025:StreamlinePerformanceAnalyzer, Google:2025:TFLiteProfiler, Qualcomm:2024:SnapdragonProfiler}} &
\textbf{Campaign Diagrams (This Work)} \\
\midrule

When It Exists &
Design-time &
Design-time &
After execution &
Design-time \& post-simulation \\
  \arrayrulecolor{verylightgray}\midrule\arrayrulecolor{black}

Primary Purpose &
Utilization \& limits &
Multi-level utilization \& limits &
Identify hot spots &
Phase-aware bottleneck \& utilization analysis \\
  \arrayrulecolor{verylightgray}\midrule\arrayrulecolor{black}

Granularity &
Coarse &
Coarse &
Fine &
Medium (Einsum-level) \\
  \arrayrulecolor{verylightgray}\midrule\arrayrulecolor{black}

Time-Aware &
No (aggregate) &
No (aggregate) &
Yes (measured) &
Yes (phases) \\
  \arrayrulecolor{verylightgray}\midrule\arrayrulecolor{black}

Hardware Target &
Any (abstract) &
Any (abstract) &
CPU, GPU, accelerators &
Any \\
  \arrayrulecolor{verylightgray}\midrule\arrayrulecolor{black}

Workload Scope &
Single Workload &
Single Workload &
Compiled, ML workloads &
Multi-phase workloads \\
  \arrayrulecolor{verylightgray}\midrule\arrayrulecolor{black}

Arch.\ modeling \& hypothetical designs &
Yes &
Yes &
No &
Yes \\
\bottomrule
\end{tabulary}
\end{scriptsize}
\end{table*}

\subsubsection*{Runtime Profilers and Execution Traces}
Runtime profilers~\cite{MediaTek:2024:NeuroPilotSDK, Arm:2025:StreamlinePerformanceAnalyzer, Google:2025:TFLiteProfiler, Qualcomm:2024:SnapdragonProfiler} provide a complementary perspective by grouping performance into specific functions or code regions. These tools are used to identify hot spots, reporting metrics such as the percentage of total runtime spent in a given function.
Additionally, execution traces~\cite{Google:2025:PerfettoTracingTool, Sun:2021:Daisen, snider:2023:hotline} report events at each time point.
While useful for debugging, profiling provides limited insight into \emph{why} a region is expensive.
Profilers typically do not distinguish whether a phase is compute-bound or memory-bound, nor do they expose how resource utilization changes across execution.
For example, in a given ML workload, a single logical computation may be realized as multiple function calls. In such cases, profiling fragments coherent algorithmic phases across multiple entries, obscuring the higher-level algorithmic structure and complicating comparison across implementations (since different implementations may use different functions for the same phase of computation).
Detailed execution timelines further exacerbate this issue: while precise, they occur after execution, are difficult to compare, and are poorly suited for design-time reasoning.

\subsubsection*{Roofline Models}\label{sssec:rooflines} A roofline model~\cite{williams:2009:RIV} is an insightful tool for determining the bottlenecks and resource limitations of a given workload on a particular architecture.
The operational intensity, OI, determines whether a workload is limited (``bound'') by memory bandwidth or compute throughput.
In an ideal world, a given workload would keep all resources busy, achieving full memory bandwidth utilization and full compute throughput utilization.
Given a roofline plot, one can then adjust various aspects of the underlying workload to improve either operational intensity or compute throughput.

\begin{figure}
  \centering
  \begin{subfigure}[t]{0.9\columnwidth}
    \centering
    \includegraphics[width=\linewidth]{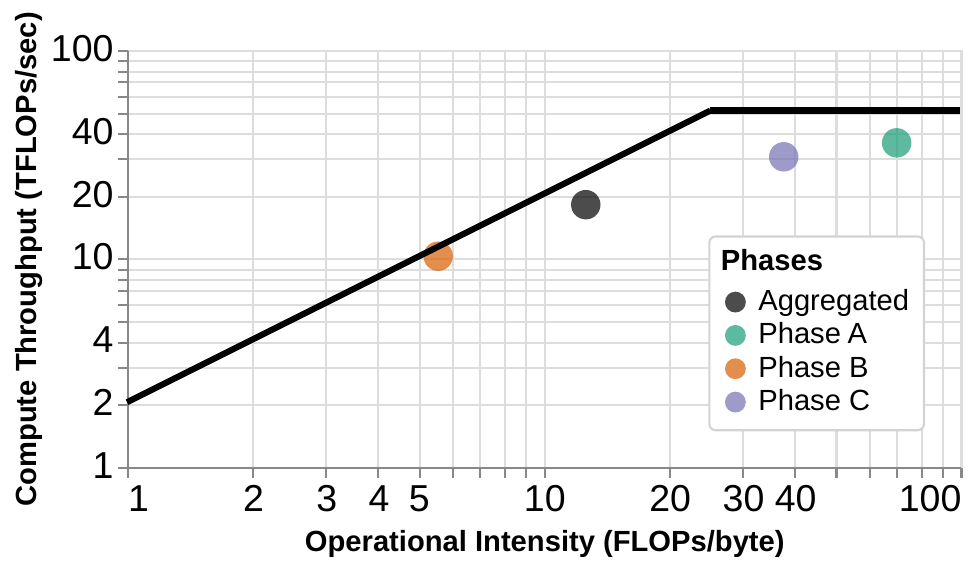}
    \caption{Roofline visualization.}\label{fig:3-kernel-roofline}
  \end{subfigure}
  \begin{subfigure}[t]{0.9\columnwidth}
    \centering
    \includegraphics[width=\linewidth]{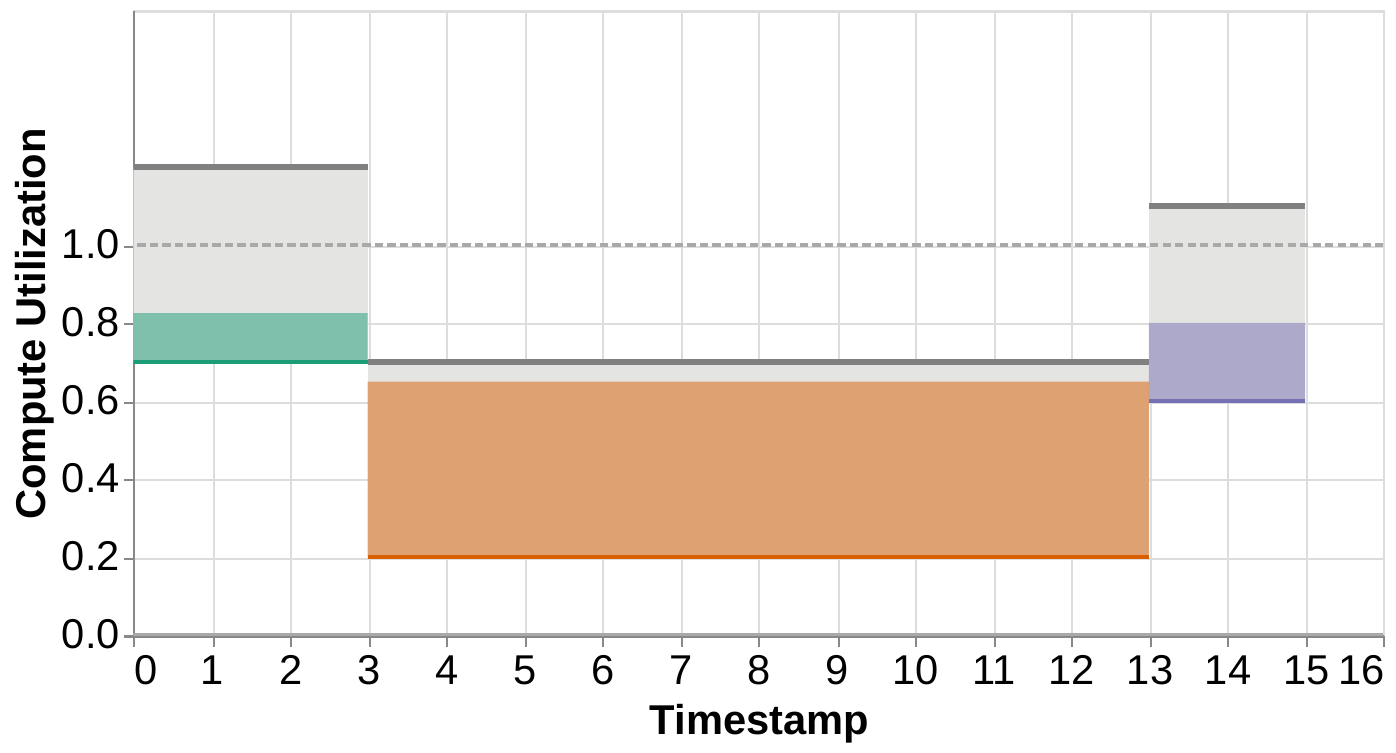}
    \caption{Campaign diagram for the same workload.}\label{fig:3-kernel-campaign}
  \end{subfigure}

  \caption{(a) Comparison of a roofline model (left, black dot), per-phase roofline (colored dots), and (b) a campaign diagram for a simple three-phase workload.}\label{fig:3-kernel}
\end{figure}

Rooflines inherently represent an \emph{aggregation over some timescale}.
A workload may be reduced to a single operating point, implicitly assuming uniform behavior throughout execution.
This aggregation obscures temporal variation within a workload, masking how resource utilization shifts across different phases.
For example, Figure~\ref{fig:3-kernel-roofline}, when looking at the \emph{black dot only}, shows the roofline for a sample workload.
The figure shows the overall workload is memory-bound, but it is difficult to determine, from the figure alone, which phase of the workload is responsible.

Attempts to refine this view by increasing temporal granularity introduce additional challenges: the execution time of each phase is unclear, and the visualization may become difficult to read as several points begin to overlap.
As a result, rooflines are poorly suited for reasoning about how performance evolves over time or how individual phases contribute to overall execution behavior.

\paragraph{What if we decomposed rooflines into phases?}
\begin{table}
\caption{What do rooflines, per-phase rooflines, and campaign diagrams reveal about a multi-phase workload? \harveyBallFull{} = fully revealed; \harveyBallHalf{} = partially revealed; \harveyBallNone{} = not revealed.}
\label{tab:vs-rooflines}
\centering
\scriptsize
\setlength{\tabcolsep}{2pt}
\renewcommand{\arraystretch}{1.0}
\vspace{-5pt}
\begin{tabular}{@{}p{0.46\columnwidth}ccc@{}}
\toprule
 \textbf{Reveals\ldots} & \textbf{Aggregate} & \textbf{Per-Phase} & \textbf{Campaign} \\
 & \textbf{Roofline} & \textbf{Roofline} & \textbf{Diagram} \\
\midrule
Bounding resource (overall)              & \harveyBallFull{} & \harveyBallFull{} & \harveyBallFull{} \\
Bounding resource (per phase)            & \harveyBallNone{} & \harveyBallFull{} & \harveyBallFull{} \\
Phase runtime                            & \harveyBallNone{} & \harveyBallNone{} & \harveyBallFull{} \\
Compute/memory traffic volume per phase  & \harveyBallNone{} & \harveyBallNone{} & \harveyBallFull{} \\
Pipelining opportunities                & \harveyBallNone{} & \harveyBallNone{} & \harveyBallFull{} \\
Dilation / throttling trade-offs         & \harveyBallNone{} & \harveyBallNone{} & \harveyBallFull{} \\
Oversubscribed resources (concurrent execution)  & \harveyBallNone{} & \harveyBallNone{} & \harveyBallFull{} \\
\bottomrule
\end{tabular}
\end{table}

To better understand the workload's behavior, we may decide to then plot a separate point for each \emph{phase} of the workload.
For our sample workload, these are the colored dots in Figure~\ref{fig:3-kernel-roofline}.
Despite these points, it is not clear \emph{which phase} is the bottleneck.
Although phase B is memory bandwidth-bound, either phase A or phase C could potentially be the bottleneck: the roofline does not tell us \emph{for how long} each of phase A and phase C stays at that level of compute throughput.
Conversely, merging multiple phases into a single dot will likely result in the dot not being on the roofline.
Since it is now a single dot, it is unclear \emph{why} it does not hit the roof.

Table~\ref{tab:vs-rooflines} lists some of the attributes missing in roofline plots that we would like campaign diagrams to support.
Overall, per-phase roofline plots indicate which resource bounds each phase (compute or memory), but they still do not indicate the (1) runtime of each phase and (2) compute and memory volumes. These attributes are important in understanding when fusion and concurrent execution of phases (or pipelining) may be helpful, among other optimizations.
If trying to model concurrent execution, rooflines may not always flag oversubscribed resources (when one or more phases attempt to use more than 100\% of a resource), since the total \emph{average} may still be lower than the roofline limits.
Meanwhile, a campaign diagram (1) visually shows the order of phases; (2) depicts how much of a resource is underutilized even when one or more phases are running; (3) flags oversubscription during concurrent execution; and (4) when stacked, the analytical versus achieved diagrams clearly show where execution deviates from the speed-of-light (\S~\ref{sec:case-mamba}).

\paragraph{Summary}
Existing approaches have a clear trade-off.
Rooflines provide global, hardware-aware insight but lack temporal and phase-level resolution, while profilers and execution traces provide detailed, post-execution data but obscure the underlying algorithmic structure.
The gap is a representation that preserves execution order, how a workload evolves with time, and resource utilization while remaining abstract enough to support various workloads and target various hardware backends.

\section{Introduction to Campaign Diagrams}\label{sec:simple-campaign}
A \emph{campaign diagram} is our proposed visual performance model that incorporates time into performance modeling analysis.
We are inspired by Minard's famous statistical graphic of Napoleon's march~\cite{tufte:1983:VDQ}, which visualizes Napoleon's Russian campaign (Figure~\ref{fig:march}).
It successfully uses a single 2D graph to display six characteristics: the (1) size of the army (width of lines) at a (2) particular location at a (3) given time, along a (4) particular direction with the (5) temperature and (6) geographical features.
Likewise, campaign diagrams encode six characteristics: (1) memory bandwidth utilization and (2) compute throughput utilization over (3) time, along with (4) compute volume, (5) memory traffic volume, and (6) phase ordering.

Intuitively, campaign diagrams show workload utilization as a ``pipe'' where both the position of the pipe and the fullness of the pipe correspond to resource utilization as it changes over time.
Figure \ref{fig:3-kernel-campaign} shows the same workload phases from the previously introduced roofline, now presented as a campaign diagram. We do not expect the reader to fully understand all details right away as campaign diagrams are informationally dense. For now, note simply that the x-axis has changed to time, clearly demonstrating phase-change behavior. Additionally, note the relative time-widths of the phases, and how that information is not communicated via the roofline which focuses on the phase rates, but not their relative weightings.\footnote{We originally tried addressing this by changing the size of the markers on the roofline to represent relative volume, but found this unsatisfying. Larger markers obscured each other easily, and the ordering of phases was visually lost. Our dissatisfaction with this solution led to the development of campaigns.}

In the remainder of the section we walk the reader through the visual language of campaign diagrams needed to fully understand this diagram, using annotated diagrams that carefully explain the intended information content. These extra annotations are not intended to be present in final campaign diagrams used in practice, which we expect to ultimately resemble Figure \ref{fig:3-kernel-campaign}.

\subsection{Anatomy of a Campaign: A Single Phase}
\begin{figure}
  \centering
  \includegraphics[width=.9\columnwidth]{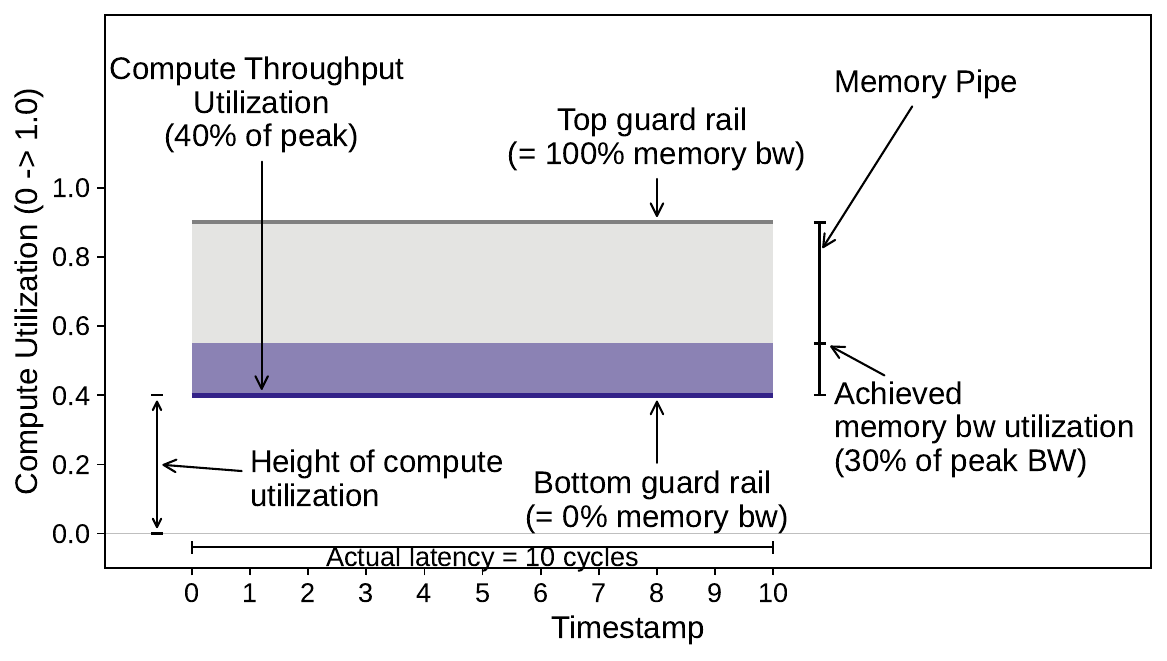}
  \caption{Anatomy of a single-phase campaign diagram. For clarity, we focus on memory bandwidth and compute throughput, but this visualization technique supports any pair of rate-limited resources.}
  \label{fig:1-phase}
\end{figure}

\begin{figure}
  \centering
  \includegraphics[width=.9\columnwidth]{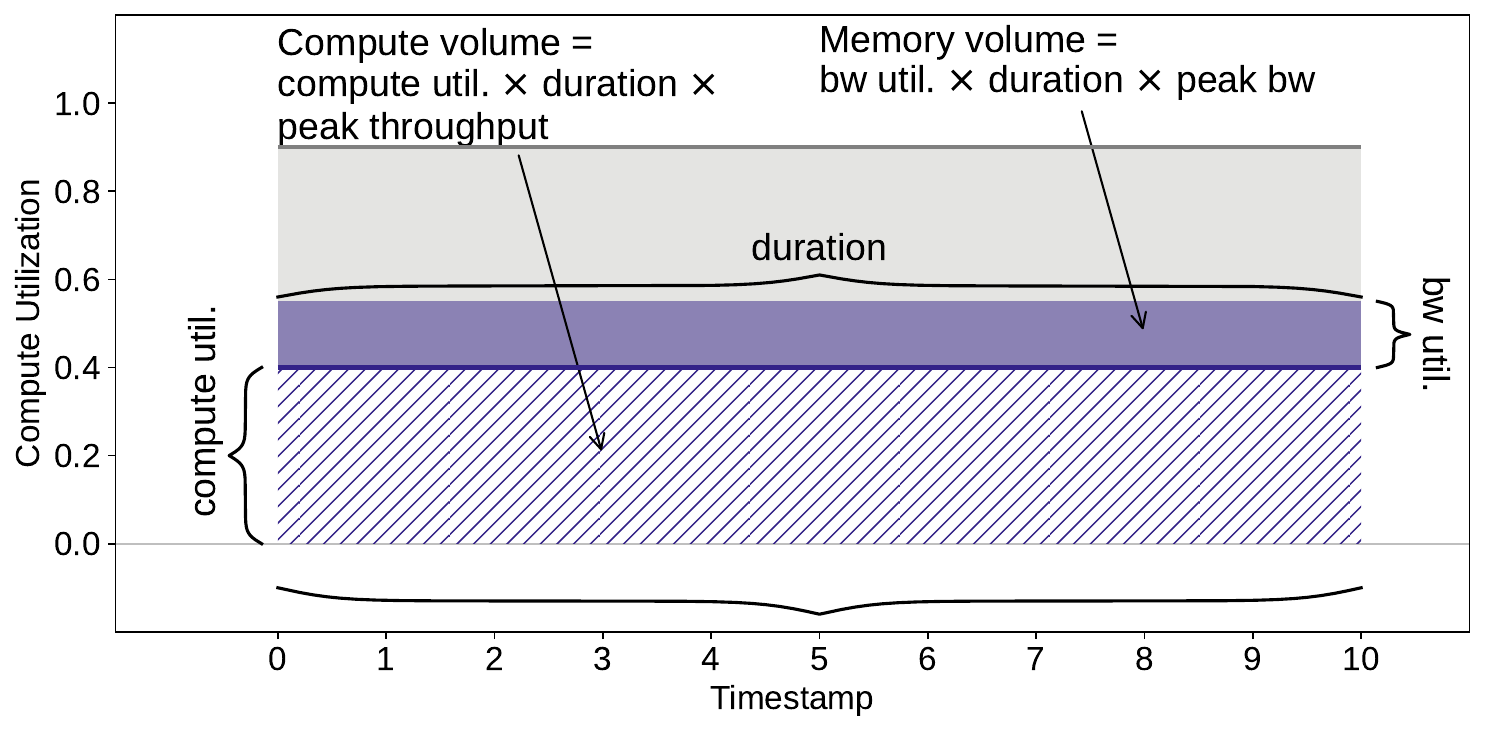}
\caption{Resource volumes: compute volume (area under the compute line) and memory traffic volume (shaded region within the pipe).}
  \label{fig:volumes}
\end{figure}

Consider the campaign diagram of a single phase in Figure~\ref{fig:1-phase}.
The time scale of the x-axis is arbitrary and it can be in whatever scale the user finds relevant (e.g., cycles, microseconds, seconds, etc.). The y-axis shows compute throughput utilization, ranging from 0 to 100\%. We find that relative utilization is most helpful, but this is not fundamental and could instead be in absolute units like TFLOPs/sec, as in rooflines. In general, relative utilization is better for workload optimization within a fixed system, and absolutes are better for cross-system comparison or design-space exploration.

The \emph{height} of the solid, colored line above the bottom of the line's position on the y-axis is the compute throughput utilization for that phase.
Each phase has a corresponding \emph{memory bandwidth utilization pipe}, which represents the total memory bandwidth available.
The pipe has two \emph{guard rails}: the compute utilization line itself is the bottom of the pipe, and the solid grey line above is the top.
We term the vertical region that the pipe spans the ``pipe width.''
The height of colored, shaded region within the pipe indicates how much of the memory bandwidth the phase uses, similar to a pipe filling with liquid.
For example, if a particular phase fully uses memory bandwidth, the area is fully colored, but if the phase only uses half the memory bandwidth, the area is half-colored from the bottom up.
The grey region is uncolored, and represents the unused memory bandwidth for a given phase. Applying a light shading helps visually distinguish the pipe from the surrounding diagram.

The absolute height of the rectangle (i.e., width of the pipe) is not significant\footnote{In this work we set each pipe to half the y-axis height; the visualized memory area is therefore $2\times$ its true magnitude relative to the compute area.}.
Note that this means that the visible area of the campaign diagram can extend above the 1.0 compute utilization top of the y-axis.
This is not because compute utilization exceeds 100\%, but simply because we need space to draw the pipe when compute utilization is maximum. We emphasize this by not drawing vertical grid lines across this region, as in Figure \ref{fig:3-kernel-campaign}.

As shown in Figure~\ref{fig:volumes}, campaign diagrams encode information beyond time-based resource utilization: they show the \emph{volume} of compute (area under the pipe) for a given phase, and the \emph{volume} of memory traffic (area of the shaded region in the pipe).
Unlike rooflines, campaign diagrams do not encode operational intensity as a first-order visual axis. However, we note that operational intensity is still present in the \emph{ratio} of compute to memory volume. Furthermore, given a precise enough diagram, the reader can visually estimate the total number of operations and/or memory transfers a phase performs. This is further information not present in traditional roofline plots.

\begin{figure}
  \centering
  \begin{subfigure}[t]{0.9\columnwidth}
    \centering
    \includegraphics[width=\linewidth]{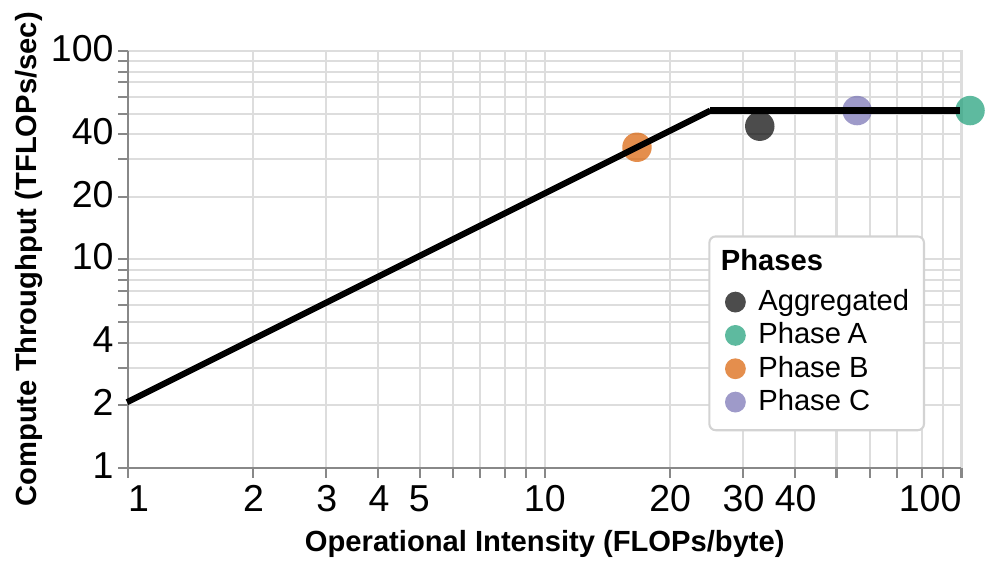}
    \caption{Roofline visualization.}\label{fig:3-phase-fused-roofline}
  \end{subfigure}\hfill
  \begin{subfigure}[t]{0.9\columnwidth}
    \centering
    \includegraphics[width=\linewidth]{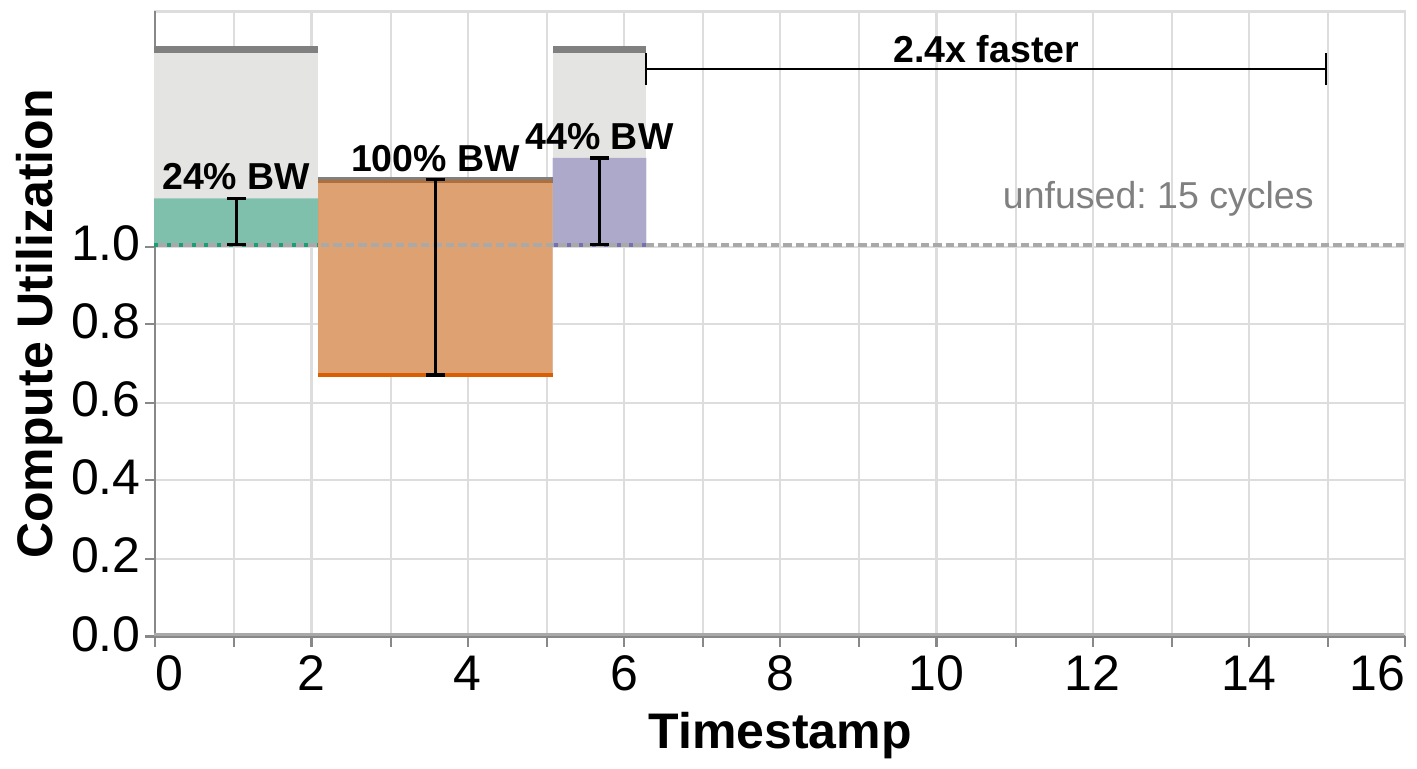}
    \caption{Fused campaign diagram.}\label{fig:3-phase-fused-campaign}
  \end{subfigure}

  \caption{The simple workload after fusion, which eliminates off-chip traffic for the intermediate tensors.}\label{fig:3-phase-fused}
\end{figure}

\subsection{Pedagogical Walk-Through}

At this point, the reader has the visual vocabulary to more deeply understand the campaign in Figure~\ref{fig:3-kernel-campaign}, and how it relates to the traditional roofline plot. For completeness, we walk through the comparison in detail, in order to make all of the informational components explicit.

\paragraph{Assumptions} For this three-phase, sample kernel, we assume that the kernel is sequential: the results of phase A are needed by phase B and the results of phase B are needed by phase C.
We assume two types of resources: main memory (either HBM or DRAM) and compute units.
Finally, to generate this diagram, assume a previous step collected (1)~memory bandwidth numbers for each phase of the kernel, (2)~compute throughput for each phase, and (3)~the runtime of each phase.

From the diagram, we ascertain that phase A has a compute throughput utilization of 70\% (0.7), phase B has a utilization of 20\% (0.2), and phase C has a utilization of 60\% (0.6) (heights of colored, solid lines).
If the underlying architecture is a Hopper GPU, which has a peak non-tensor-core FP32 throughput of 51.2~TFLOPs/sec~\cite{NVIDIA:2023:H100}, and the timescale is in seconds, then the workload has throughputs of 35.8~TFLOPs/sec, 10.2~TFLOPs/sec, and 30.7~TFLOPs/sec for each phase, respectively.
Thus, we can see that phase A uses 70\% of the compute resources for a quick, 3 second burst, phase B uses 20\% of the compute resources for 10 seconds, and phase C uses 60\% of the compute resources for a 2-second burst.\footnote{Total work on Hopper: 107.4, 102, and 61.4~TFLOPs for phases A, B, and C, respectively.}

In our example, phase A ``fills up'' 25\% of its memory pipe (25\% of its area is colored); it uses 25\% of the memory bandwidth. Likewise, phases B and C use 90\% and 40\% of the memory bandwidth, respectively. The memory volume of each phase (in bytes) corresponds to the area of its colored region.

\emph{Insights from the Diagram.}
We can now draw conclusions at a glance. In our example, since phase B is the widest rectangle, we know it takes the largest fraction of runtime (thus, a good place to start for optimization). Since no compute line is fully solid, we also know that no phase is fully utilizing the underlying architecture's compute capability.
Impactful optimizations must target reducing the overall memory traffic (or modifying the architecture to increase memory bandwidth), and finding ways to improve overall resource utilization (fusion, various mappings and schedules, algorithmic changes).

\subsection{Visualizing Cross-Phase Fusion}\label{ssec:simple-fusion}
Suppose as domain experts, we know that phase C reads the output of phase B, and phase B reads the output of phase A.
Then, there is an opportunity for \emph{fusion}, where each phase keeps its output on-chip.
By keeping intermediate data on-chip, fusion eliminates the memory traffic between producer and consumer phases that would otherwise have gone to main memory.
The corresponding roofline plot (Figure~\ref{fig:3-phase-fused-roofline}) shows the aggregate and per-phase points.
Figure~\ref{fig:3-phase-fused-campaign} shows the corresponding campaign diagram.
Phase B, previously the bottleneck, now fully achieves 100\% memory bandwidth utilization, as fusion reduced its memory volume (colored region) over a shorter duration.
Similarly, phases A and C, due to less memory volume, are able to fully utilize the compute resources (solid, colored line at 100\% compute utilization). The overall runtime improves by 2.4$\times$.
The campaign diagram makes it clear how much each phase contributes to the overall runtime, as well as the overall change in compute and memory volume compared to the unfused diagram.
We demonstrate this benefit on a real workload in \S~\ref{sec:case-mamba}.

\paragraph{Summary}
Campaign diagrams enable a clear understanding of how different optimization decisions impact resource utilization (and performance), as well as provide insight into certain types of optimization opportunities.
Workloads are quite complex, and in the following sections we illustrate an increasingly complex set of workloads that mimic the characteristics of modern workloads.
Many of the optimizations we walk through are already popular and empirically beneficial optimizations; this paper is not meant to advocate for them but instead to use them as examples of complex but helpful optimizations that are difficult to identify.
Existing visualization tools do not allow us to identify opportunities for these optimizations.

\section{Optimizing Using Campaign Diagrams}\label{sec:pipeline}
We now show how campaign diagrams expose optimization opportunities, the oversubscription of resources, and the impact of scheduling changes on a kernel, particularly for \emph{pipelining}.
Pipelining allows multiple phases of computation to run concurrently; for example, memory access for one phase may occur while compute for a different phase executes~\cite{lam:1988:spe} or different compute units perform different tasks~\cite{Nayak:2024:FML_micro}.
To leverage pipelining, the underlying hardware and programming model must be able to support concurrent execution of different tasks.

\subsection{Optimizing a 2-Phase Sample Workload}

\begin{figure}
  \centering
  \begin{subfigure}[t]{0.9\columnwidth}
    \centering
    \includegraphics[width=\linewidth]{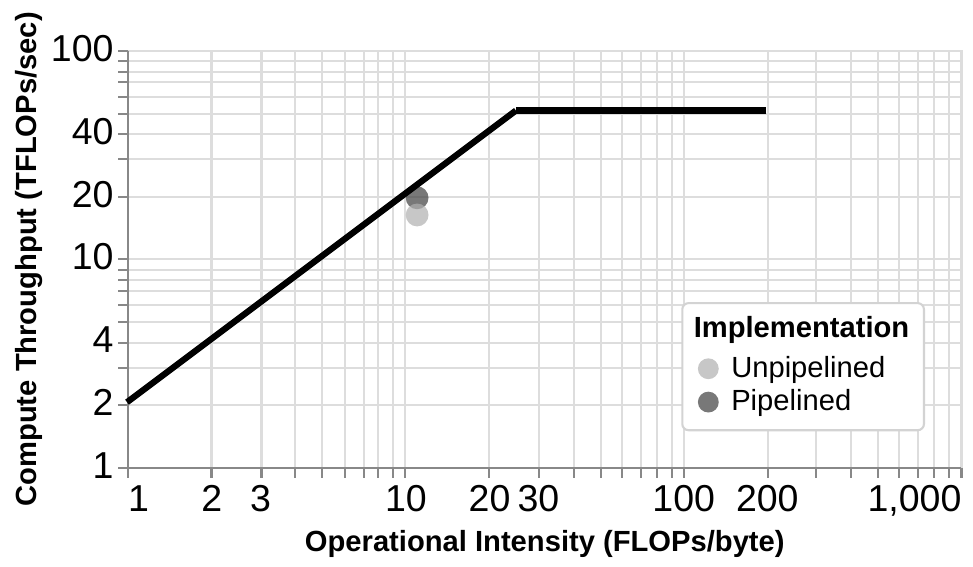}
    \caption{Aggregated points for the unpipelined and pipelined implementations.}\label{fig:2-kernel-roofline}
  \end{subfigure}
  \begin{subfigure}[t]{0.9\columnwidth}
    \centering
    \includegraphics[width=\linewidth]{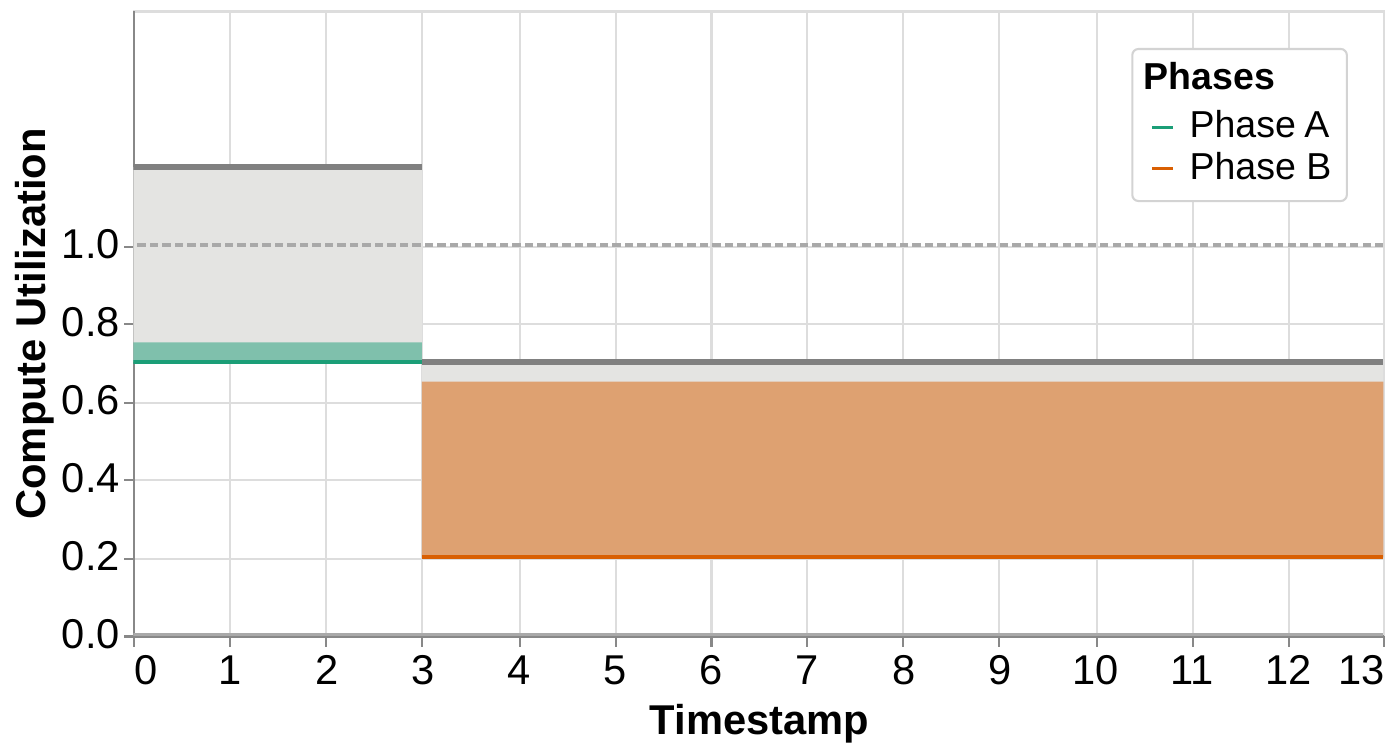}
    \caption{Campaign diagram for the same workload.}\label{fig:2-kernel-campaign}
  \end{subfigure}
  \vspace{-5pt}
  \caption{A two-phase workload.}\label{fig:2-kernel}
\end{figure}

\begin{sidebar_box}[label=inset:2-phase]{A Sample, 2-Phase Workload}
Figure~\ref{fig:2-kernel-campaign} shows the campaign diagram for a sample workload with two phases.
The workload runs for 13 time units, with phase A running for 3 time units, and phase B for 10 time units.
Phase A has a compute throughput utilization of 70\% and memory bandwidth utilization of 10\%.
Phase B has a compute throughput utilization of 20\% and memory bandwidth utilization of 90\%.
The diagram shows that phase A is compute-bound (as the compute line is longer than the memory line),
while phase B is memory bandwidth-bound (memory line is longer than the compute line).
Figure~\ref{fig:2-kernel-roofline} is the equivalent roofline.
\end{sidebar_box}

\begin{figure}
  \centering
  \begin{subfigure}{0.9\columnwidth}
    \centering
    \includegraphics[width=\linewidth]{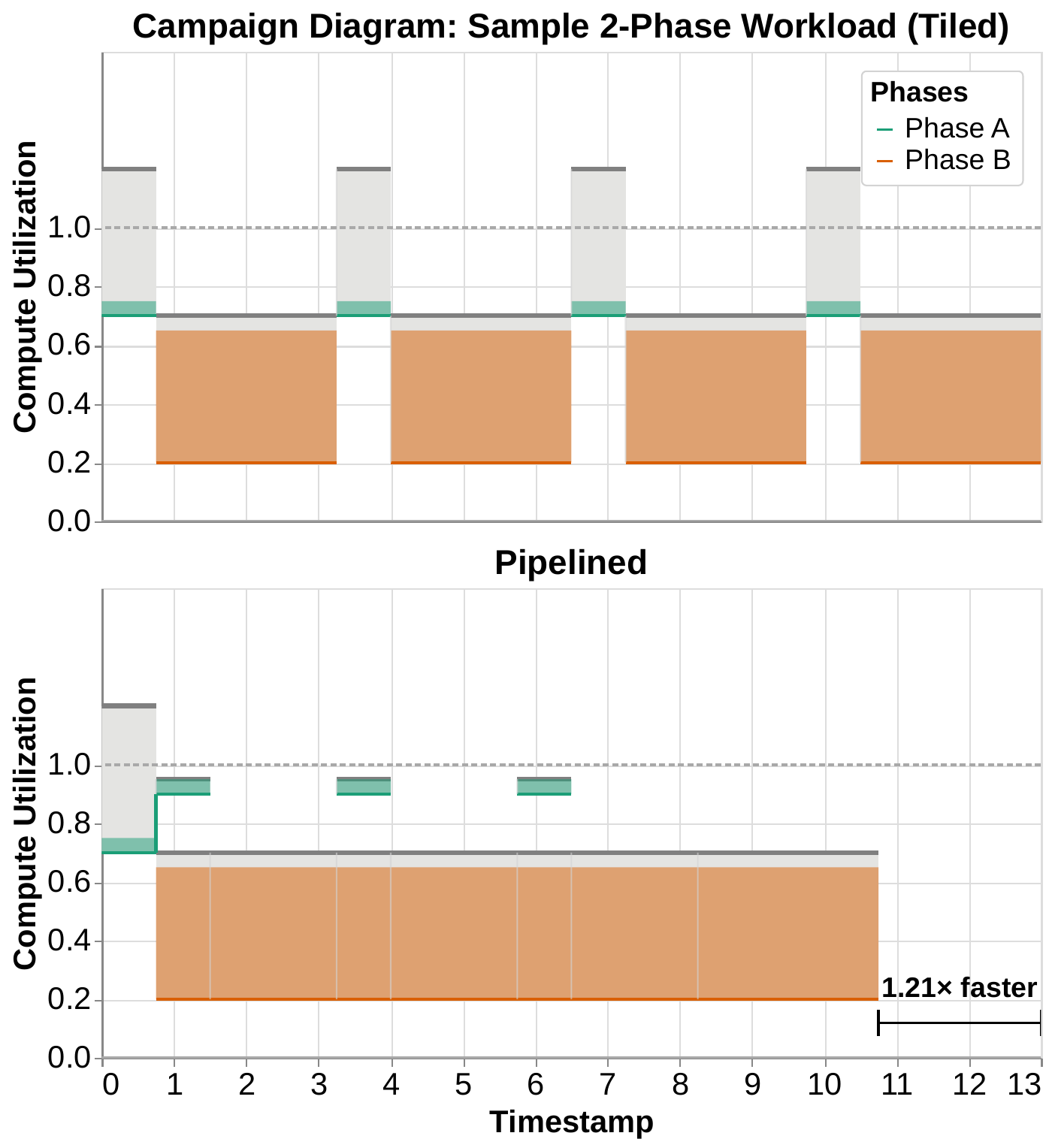}
    \caption{Tiled (top) and pipelined (bottom) on a shared time axis.}\label{fig:2-kernel-stacked}
  \end{subfigure}

  \begin{subfigure}{0.9\columnwidth}
    \centering
    \includegraphics[width=\linewidth]{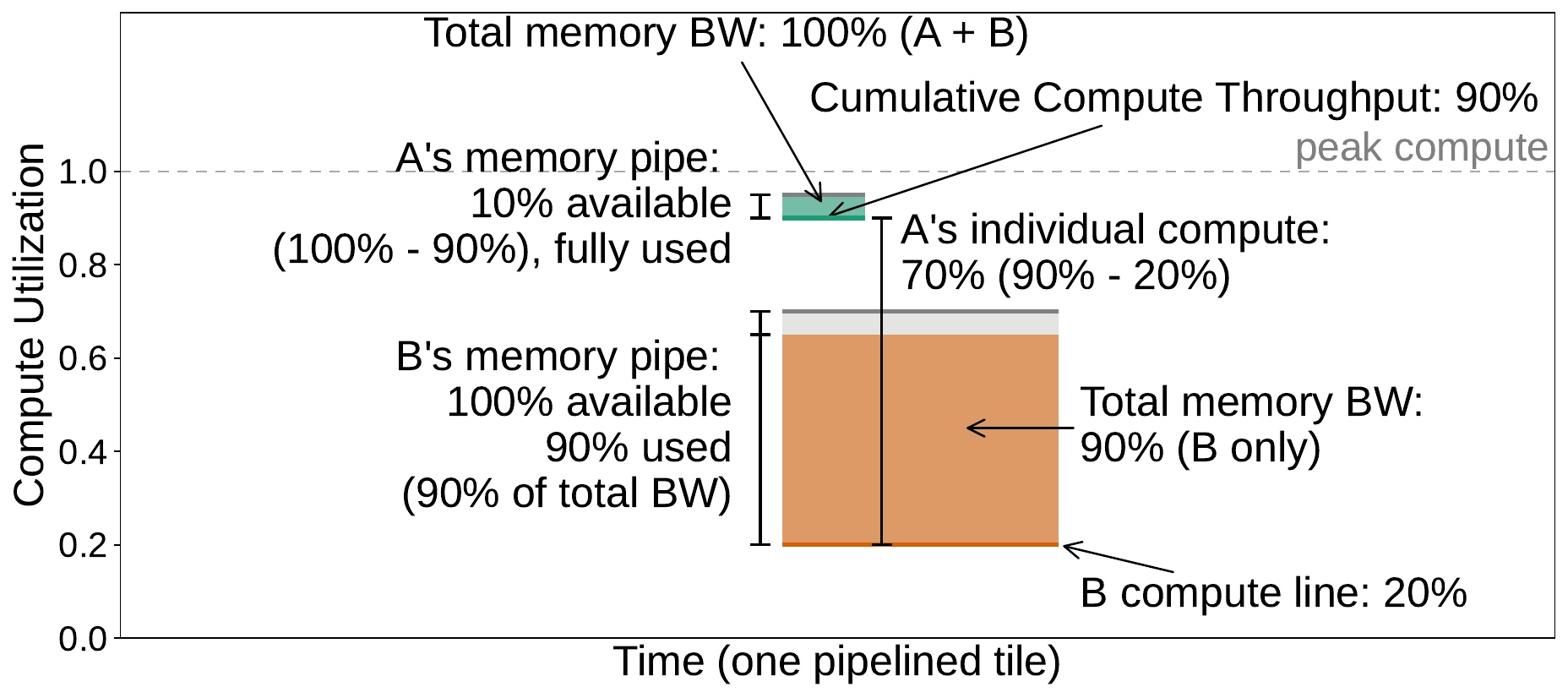}
    \caption{Zoom into one steady-state pipelined tile.}\label{fig:2-kernel-zoom}
  \end{subfigure}
  \caption{(a)~Tiling the kernel, then pipelining the tiled schedule so the phases overlap. (b)~Anatomy of concurrent execution, showing how phase~A's burst stacks atop phase~B.}\label{fig:2-kernel-opt}
\end{figure}

Campaign diagrams reveal scheduling decisions obscured by rooflines, and these details expose opportunities for concurrent execution.
Consider another 2-phase workload, shown in Figure~\ref{fig:2-kernel-campaign} (and summarized in Inset~\ref{inset:2-phase}).
In this workload, assume phase B consumes the result of phase A, i.e., phase B \emph{depends} on phase A.
Since phase B depends on phase A, let us schedule such that we \emph{partition} the workload to produce tiles of A that can then be consumed by B (Figure~\ref{fig:2-kernel-stacked}, top).
While the roofline for this partitioned mapping remains the same, the campaign diagram now clearly shows when each tile executes.
A and B tiles then alternate, with each A tile independent of the previous B tile.
The campaign diagram reveals that phase A underutilizes the memory bandwidth.
Meanwhile, phase B underutilizes the compute resources.
A low memory bandwidth, high compute throughput kernel could pair well with phase B.
Since a given tiled execution of phase B depends on a previously executed tile of phase A, we can \emph{overlap} their execution.
Figure~\ref{fig:2-kernel-stacked} (bottom) shows the corresponding campaign diagram.
Concurrent execution, or pipelining, decreases the runtime from 13 time units to 10.75 time units.
This transformation is barely visible on the roofline (Figure~\ref{fig:2-kernel-roofline}): the pipelined point sits only slightly above the unpipelined one at the same operational intensity, \emph{and} there is no equivalent ``per-phase'' roofline for concurrent execution as each phase's roofline point would remain the same as the unpipelined roofline.

\paragraph{Reading Pipelined Campaign Diagrams}
Figure~\ref{fig:2-kernel-zoom} zooms into one steady-state tile of the pipelined schedule: phase~A runs as a short burst stacked atop phase~B, so their compute lines are additive, where phase~B at the bottom contributes 20\%, and phase~A on top contributes 70\% (yielding a combined throughput of 90\%).
Each phase's individual compute utilization is the distance between its compute line and the compute line of the phase below.
Memory wells stack the same way: the bottommost phase has the full 100\% memory bandwidth well available, and each subsequent phase's well shrinks by the bandwidth utilization of the phases beneath it.
Each phase's memory bandwidth contribution is the colored, shaded region within its own well.
The choice of which phase sits at the bottom is arbitrary (placing A below and B above yields an equivalent diagram).
To know the total memory bandwidth used by all phases at a timestamp, check how full the topmost well is; the grey portion above the colored region indicates the remaining unused bandwidth. In Figure~\ref{fig:2-kernel-zoom}, the topmost bandwidth is full, hence A and B stacked achieve peak memory bandwidth utilization.

\subsection{Oversubscribing Resources}

Pipelining can reduce latency by overlapping phases with complementary resource needs.
Without careful scheduling, however, multiple phases executing at once can lead to the combined resource demand exceeding the available resource capacity.
This is resource oversubscription.
Campaign diagrams make oversubscription visible at the time and resource where it occurs.
For compute throughput, the compute lines exceed 1.0, and for memory bandwidth, phases spill beyond their memory guard rails.
Our visualization allows designers to compare candidate transformations that potentially reduce these demands (see Inset~\ref{inset:tooling}).

We introduce the terms \emph{dilation} and \emph{throttling} to describe two schedule transformations that reduce requested resource demand at a given timestamp.
\emph{Dilation}
stretches a phase's execution in time, decreasing its compute rate or memory-access rate while maintaining the same volume of compute or memory.
Campaign diagrams visualize this, but our implementation also acts as an exploratory tool: we apply a specific dilation policy that identifies the bottleneck phase, i.e., the phase with the longest execution time in a pipelined tile of execution, and stretches all other phases in time to match.
This allows a designer to quickly evaluate whether dilation effectively reduces demand.
\emph{Throttling} is then applied to the overlapped execution after dilation. It stretches all phases in a concurrent block in time until the combined requested resource demand is at or below 100\%.
In the following diagrams, the solid compute lines show the original phase duration, while dashed extensions show the additional time introduced by dilation or throttling (Figure~\ref{fig:cd-w3-anatomy}).

\begin{sidebar_box}[label=inset:tooling]{Exploring Transformations with Our Tool}
Given an upstream, unpipelined campaign diagram, our tool enables a user to see the effects of pipelining, throttling, and dilation by calling the \texttt{pipeline}, \texttt{throttle}, and \texttt{dilate} flags in our implementation.
This enables schedule exploration.
Depending on the results, a user can evaluate whether the transformation is worth implementing in their design.
An architect, for example, may need to add logic to support concurrent execution (see \S~\ref{sec:case-mamba}, also see~\cite{Nayak:2024:FML_micro}).
\end{sidebar_box}

\subsubsection{Oversubscribing Memory Bandwidth}
\begin{figure}
  \centering
  \begin{subfigure}[t]{.9\linewidth}
    \centering
    \includegraphics[width=\linewidth]{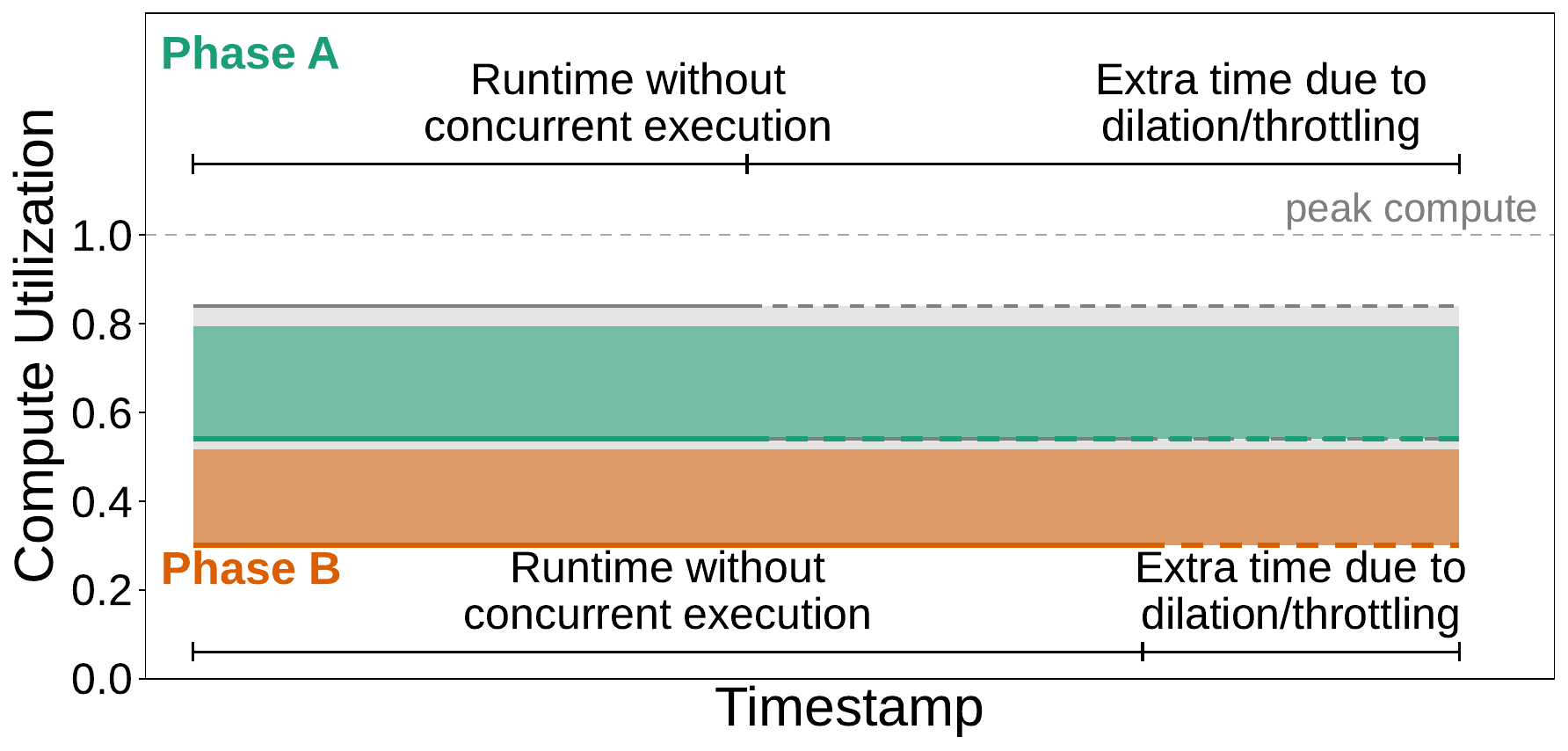}
    \caption{Anatomy of a pipelined~+~dilated~+~throttled tile.
    }\label{fig:cd-w3-anatomy}
  \end{subfigure}

  \begin{subfigure}[t]{0.9\columnwidth}
    \centering
    \includegraphics[width=\linewidth]{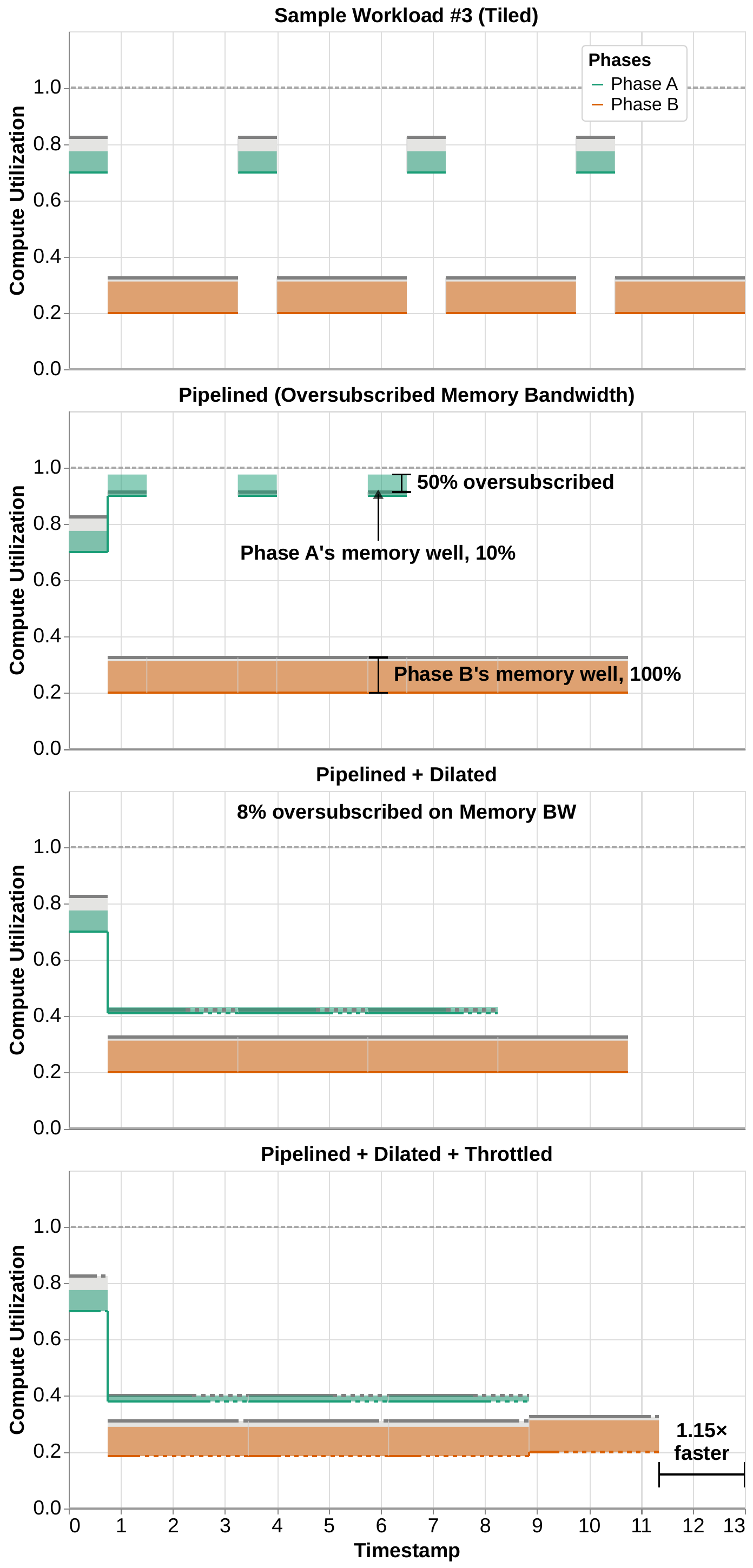}
    \caption{Memory-bandwidth oversubscription progression.}\label{fig:cd-w3-progression}
  \end{subfigure}

  \caption{Memory-bandwidth oversubscription. Campaign diagrams expose when pipelining oversubscribes resources and how dilation/throttling redistribute resource demand over time.}\label{fig:cd-w3-stack}
\end{figure}

Figure~\ref{fig:cd-w3-progression} (top panel) shows an updated sample workload, where phase A now uses 60\% of the memory bandwidth, and phase B uses 90\%.
If we attempt to pipeline this workload (Figure~\ref{fig:cd-w3-progression}, second panel), in the steady state the schedule is infeasible: the workload demands 150\% of the memory bandwidth, \emph{oversubscribing} the memory bandwidth.
Notice the colored region for phase A spills above the grey boundary line: this is memory bandwidth oversubscription.

To address this, we can \emph{dilate} phase A (Figure~\ref{fig:cd-w3-progression}, third panel) to execute in the same amount of time as the corresponding concurrent phase B block.
This reduces the memory bandwidth utilization of phase A to 18\% when pipelined with phase B.
However, even with dilation, the workload still has a memory bandwidth demand of 108\%.
To further smooth out memory bandwidth utilization, we can apply throttling to the \emph{dilated} diagram in Figure~\ref{fig:cd-w3-progression} (third panel).
The result is shown in Figure~\ref{fig:cd-w3-progression} (bottom panel): phase A produces a tile for the entirety of the length of time it takes for phase B to consume a tile.
This combined approach further reduces the runtime from 13 time units to 11.35 time units (dilate + throttle).
Overall, the campaign diagram reveals where and when oversubscription occurs, and how dilation and throttling redistribute resource demand over time.

\subsubsection{Oversubscribing Compute Throughput}
\begin{figure}
  \centering
  \includegraphics[width=0.9\columnwidth]{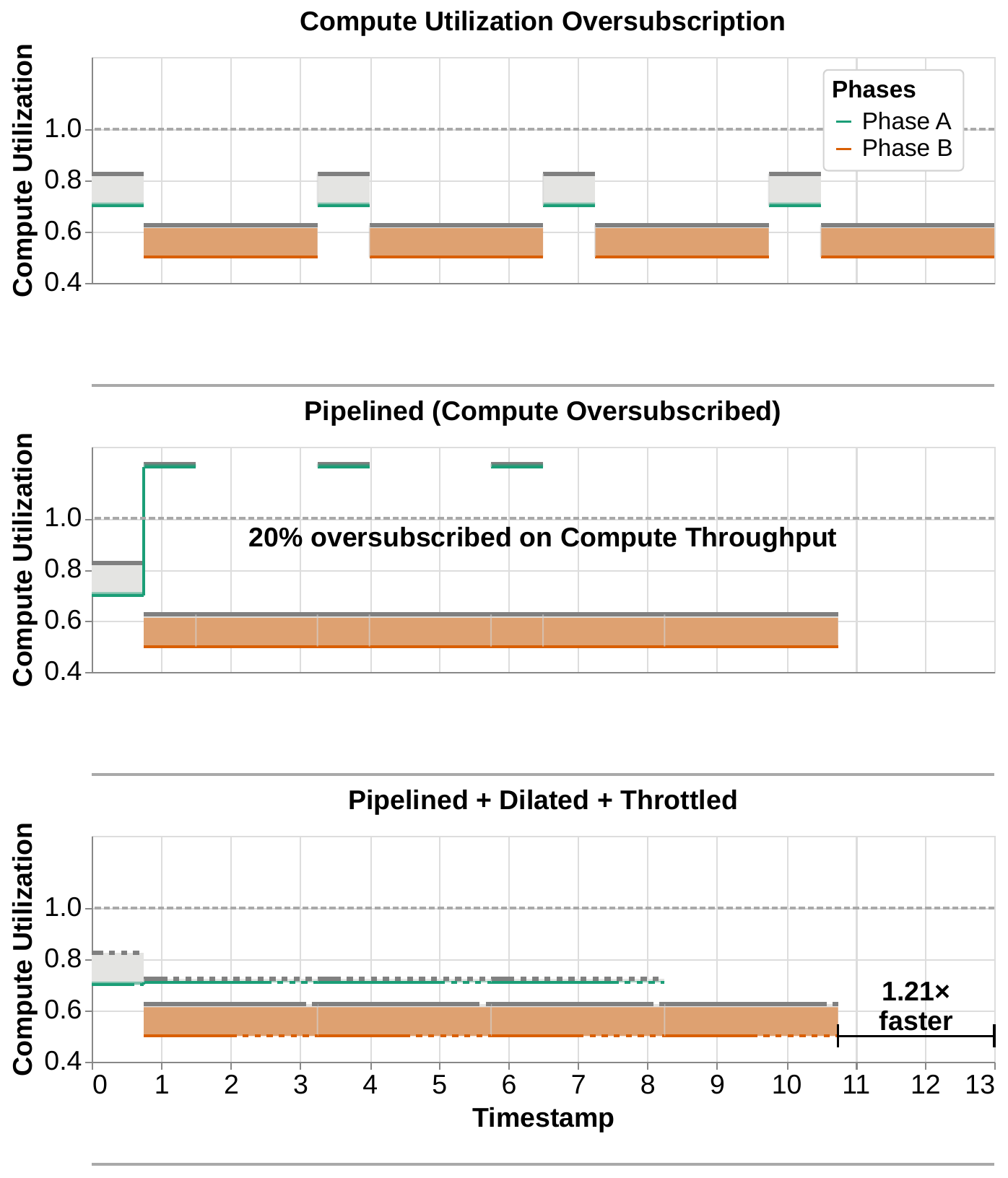}
  \caption{Oversubscribing compute throughput utilization, resolved with dilation and throttling.}
  \label{fig:4-compute-oversubscribed}
\end{figure}

The same pattern applies to compute throughput.
Figure~\ref{fig:4-compute-oversubscribed} (top) shows a modified and tiled campaign diagram that now uses 50\% compute throughput utilization in phase B.
Pipelining stacks a phase A execution tile atop phase B, leading to an oversubscription of resources at 120\% compute throughput utilization (Figure~\ref{fig:4-compute-oversubscribed}, middle).
Finally, dilating and throttling the pipelined workload (Figure~\ref{fig:4-compute-oversubscribed}, bottom) reduces the compute throughput utilization to 71\% while improving the runtime to 10.75 time units from 13 time units.

\subsubsection{Example: Optimizing a 3-Phase Workload}

\begin{figure}
  \centering
  \includegraphics[width=.9\linewidth]{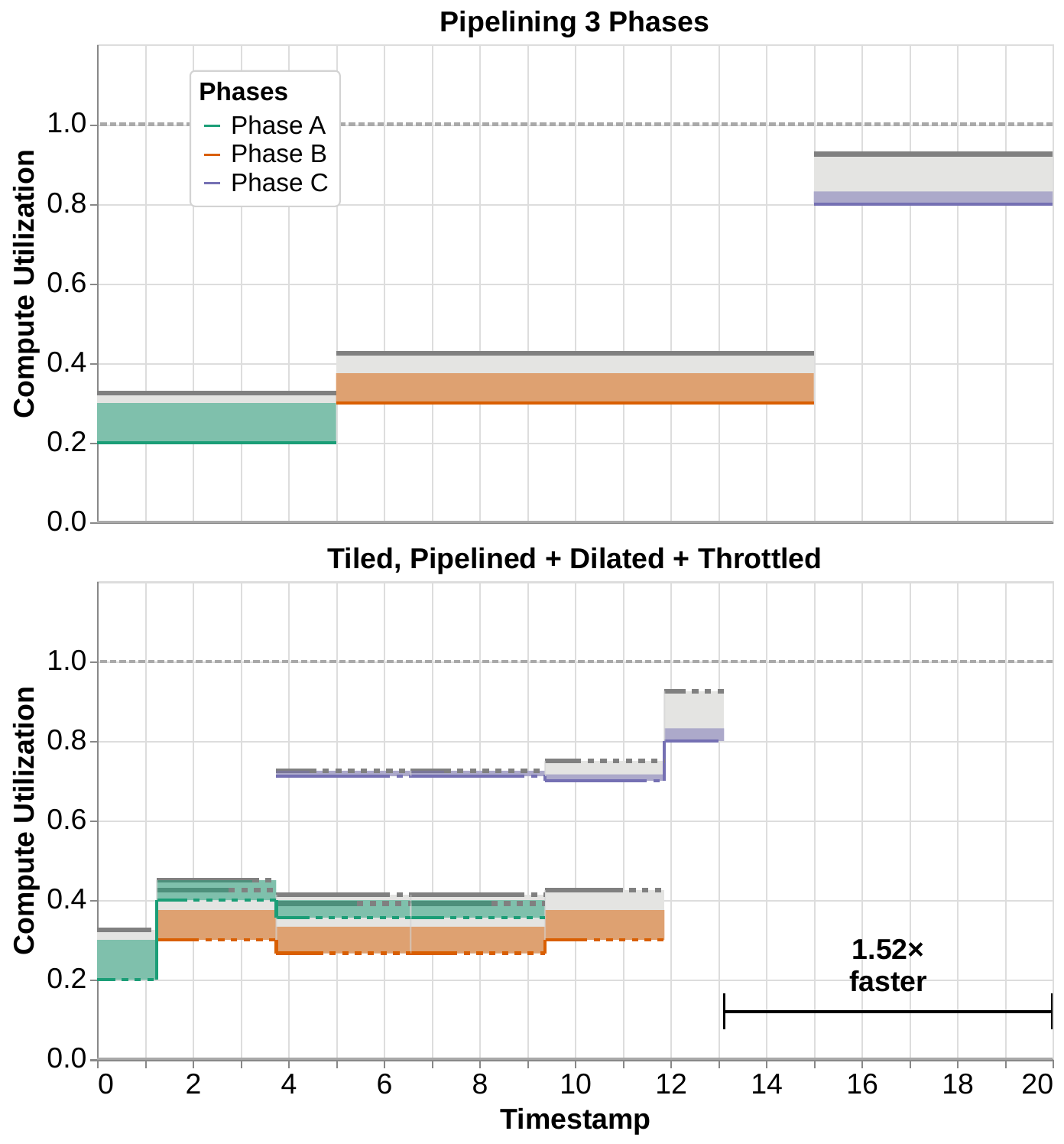}
  \caption{Pipelining on a three-phase workload (A $\rightarrow$ B $\rightarrow$ C)}
  \label{fig:3phase-campaign-roofline}
\end{figure}

The previous concepts apply to workloads with more than two phases (Figure~\ref{fig:3phase-campaign-roofline}).
Assuming that Phase~C depends on Phase~B, which in turn depends on Phase~A, we use the campaign-diagram panels to visualize the impact of tiling and pipelining.
To prevent oversubscription of both compute and memory bandwidth resources, we use the transformed campaign diagram to depict the impact of throttling and dilation.
These transformations improve the runtime from 20~time units to 13.125~time units.

\section{Case Study: Low-Rank Approximation}\label{sec:case-lr-gemm}

\emph{Modeling parameters.} Since campaign diagrams can be generated from analytical models, simulations, or measured executions, each quantitative case study depends on the hardware and traffic model used to compute phase runtimes.
Table~\ref{tab:modeling-config} summarizes the parameters used for the quantitative results in this section and Section~\ref{sec:case-mamba}.

\begin{table}
\centering
\scriptsize
\setlength{\tabcolsep}{2pt}
\renewcommand{\arraystretch}{0.95}
\caption{Hardware and modeling parameters used in the quantitative case studies.}
\label{tab:modeling-config}
\vspace{-4pt}
\begin{tabular}{@{}p{0.23\columnwidth}p{0.32\columnwidth}p{0.36\columnwidth}@{}}
\toprule
Parameter & LR-GEMM & Mamba-1 (370m) \\
\midrule
Architecture & FuseMax~\cite{Nayak:2024:FML_micro} & FuseMax-like \\
Compute layout & $256{\times}256$ 2D & $256{\times}256$ 2D $+$ 1D \\
Peak mem.\ BW & $400$\,GB/s & $2039$\,GB/s \\
Clock & $940$\,MHz & $1.75$\,GHz \\
Workload config & $N{=}20480$, $R{=}Q{=}512$ &  batch size $=64$, sequence length ${=}16384$ \\
Traffic model & alg.\ min.\ traffic & analytical best-case + Timeloop-based sim.~\cite{Parashar:2019:TSA} \\
\bottomrule
\end{tabular}
\vspace{-6pt}
\end{table}

\begin{figure}
  \centering

  \begin{subfigure}{0.9\columnwidth}
    \centering
    \includegraphics[width=\linewidth]{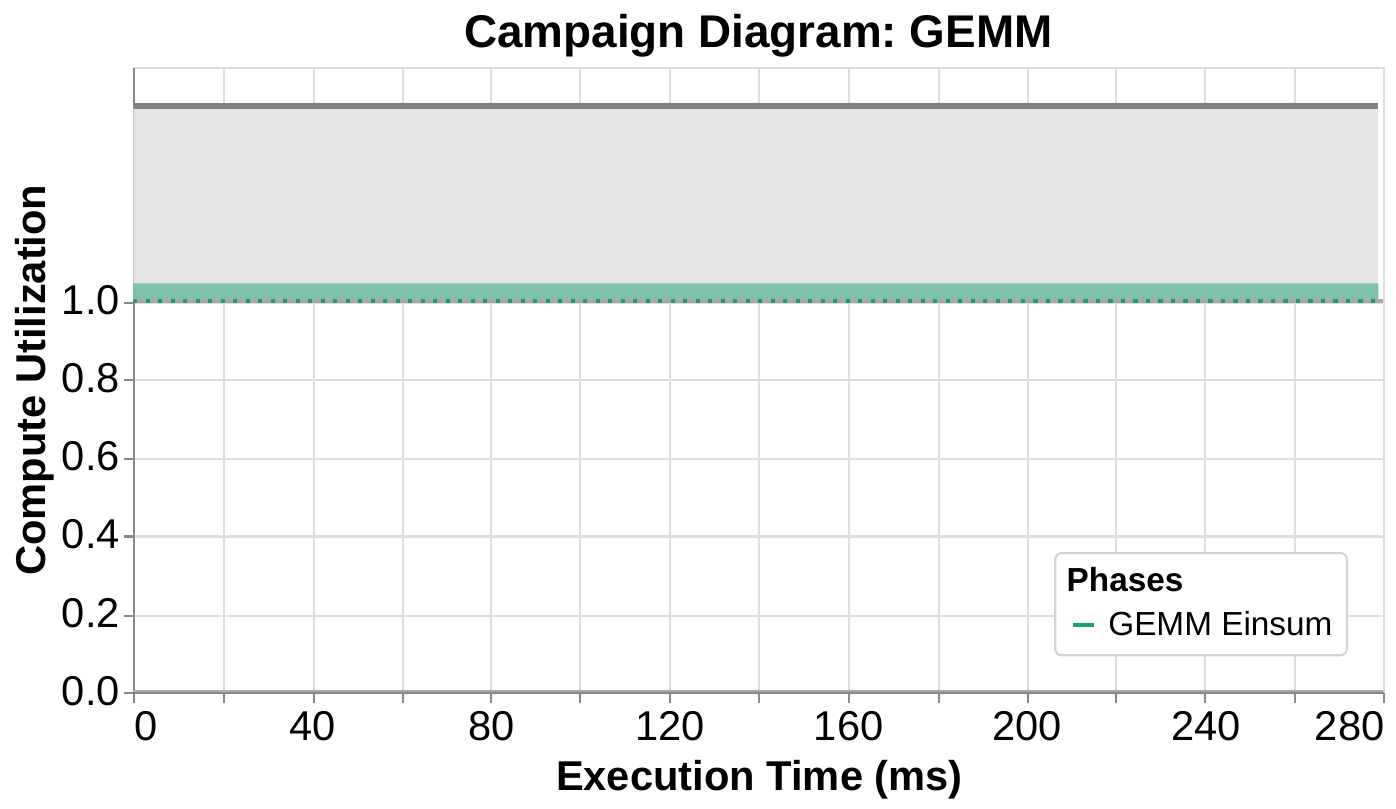}
    \caption{Dense GEMM campaign diagram.}
    \label{fig:lrgemm-dense-campaign}
  \end{subfigure}

  \begin{subfigure}{0.9\columnwidth}
    \centering
    \includegraphics[width=\linewidth]{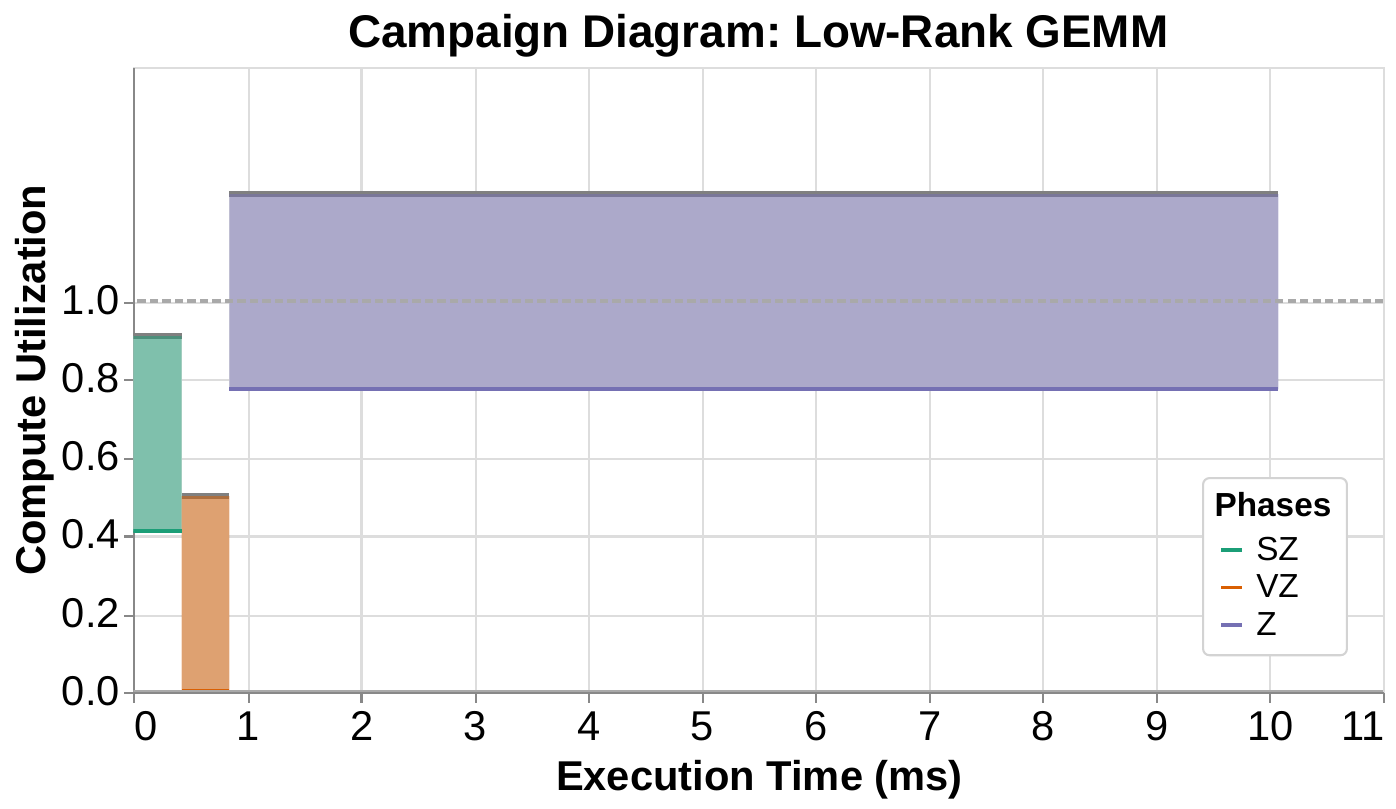}
    \caption{LR GEMM campaign diagram (decomposed).}
    \label{fig:lrgemm-decomposed-campaign}
  \end{subfigure}

  \begin{subfigure}{0.9\columnwidth}
    \centering
    \includegraphics[width=\linewidth]{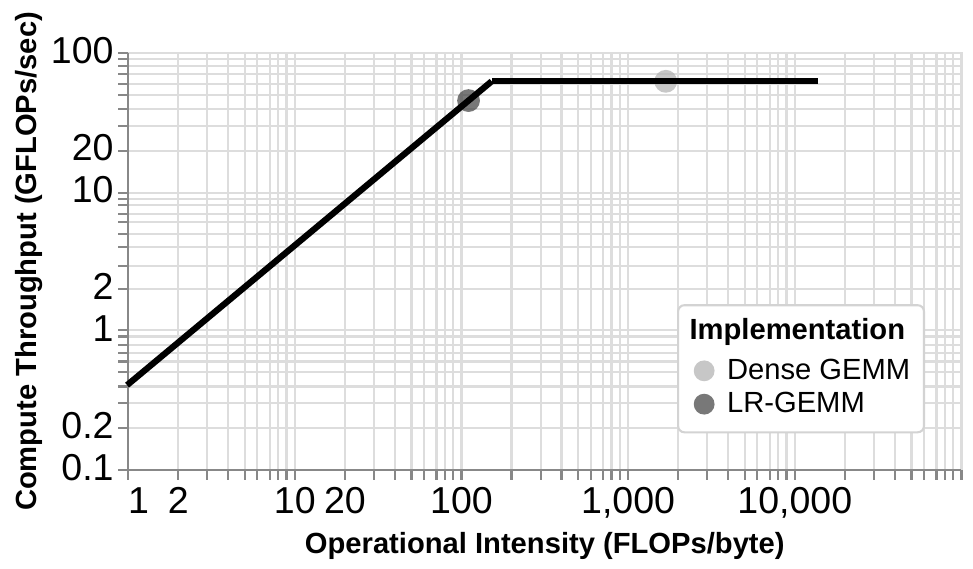}
    \caption{Aggregated roofline points for dense and LR GEMM.}
    \label{fig:lrgemm-roofline}
  \end{subfigure}

  \caption{\emph{Top:} dense campaign diagram;
  \emph{middle:} decomposed low-rank; \emph{bottom:} aggregated roofline points
  for both~\cite{metere:2025:lrg}.}
  \label{fig:lrgemm-roofline-campaign-comparison}
\end{figure}

Campaign diagrams also enable us to reason about the performance impact of
\emph{algorithmic} changes.
In this case study, we use campaign diagrams to visualize the effect of
low-rank approximation on dense matrix multiplication (GEMM), given by $Z_{m,n} =  A_{m,k} \times B_{k,n}$.

For large matrices, low-rank factorizations can reduce storage requirements, memory traffic, and computational complexity.
This technique has been applied in several domains,
including large language models and machine learning~\cite{hu:2021:lora, cheng:2020:deepeye, liu:2023:tdm}.

\paragraph{Low-rank approximation via SVD}
We approximate the input matrices using singular value decomposition (SVD):
\begin{align}
A_{m,k} &= \bigl( UA_{m,r} \times SA_{r} \bigr) \times VA_{r,k}, \\
B_{k,n} &= \bigl( UB_{k,q} \times SB_{q} \bigr) \times VB_{q,n}.
\end{align}

Here, $UA$, $SA$, and $VA$ are the SVD factors of $A$ truncated to rank $R$,
and $UB$, $SB$, and $VB$ are the corresponding SVD factors of $B$ truncated
to rank $Q$. We assume $R \equiv Q \ll \min(M, K, N)$.

\emph{Low-rank (LR) GEMM formulation.} Following Metere~\cite{metere:2025:lrg}, we use a singular value decomposition (SVD)-based low-rank (LR) formulation for GEMM:

\begin{subequations}\label{eq:lrgemm}
\begin{align}
SZ_{r,q} &= SA_{r} \times \bigl( VA_{r,k} \times UB_{k,q} \bigr), \label{eq:lrgemm-sz}\\
VZ_{q,n} &= SB_{q} \times VB_{q,n}, \label{eq:lrgemm-vz}\\
Z_{m,n} &= \bigl( UA_{m,r} \times SZ_{r,q} \bigr) \times VZ_{q,n}, \label{eq:lrgemm-z}
\end{align}
\end{subequations}
where $UA$, $SZ$, and $VZ$ serve as the SVD factors of $Z$.
Reconstructing the full output matrix requires evaluating the two GEMMs in Eq.~\eqref{eq:lrgemm-z}.

\emph{Modeling assumptions.} We adopt the same matrix dimensions as Metere~\cite{metere:2025:lrg}
($N = 20480$, $R = Q = 512$) and assume execution on the 2D array architecture
of FuseMax~\cite{Nayak:2024:FML_micro}.\footnote{FuseMax is representative of the common TPU-like accelerator pattern used for machine learning workloads.}
We further assume that the SVD has been performed \emph{offline}, and that
all SVD factors are available prior to the LR GEMM computation.

\emph{Visualization overview.} Figure~\ref{fig:lrgemm-roofline-campaign-comparison}
shows a single roofline (bottom) with the aggregated points for the dense and LR GEMM computations.
The dense GEMM is compute-bound (colored line at 1.0), yet has a long runtime (278 ms) as it has a large amount of compute.
Approximating the input matrices using SVD reduces the overall number of computations \emph{and} significantly reduces the overall runtime.
Specifically, note that the low-rank approximation transforms the previously compute-bound kernel to a memory-bound kernel.
Traditionally, workload designers often try to \emph{increase} operational intensity, since higher OI indicates more work per byte transferred.
This in turn pushes a workload toward the compute-bound region.
Indeed, by observing the roofline points for the dense GEMM and the low-rank approximation, one might believe the dense GEMM performs better.
However, the campaign diagrams reveal a behavior obscured by the roofline model: for this problem instance, the memory-bound algorithm improves performance, \emph{despite having lower operational intensity}.
It significantly reduces latency by reducing the \emph{volume} of computation and memory traffic.

\section{Case Study: Mamba State Space Model} \label{sec:case-mamba}
\begin{figure}
  \centering
  \includegraphics[width=\columnwidth]{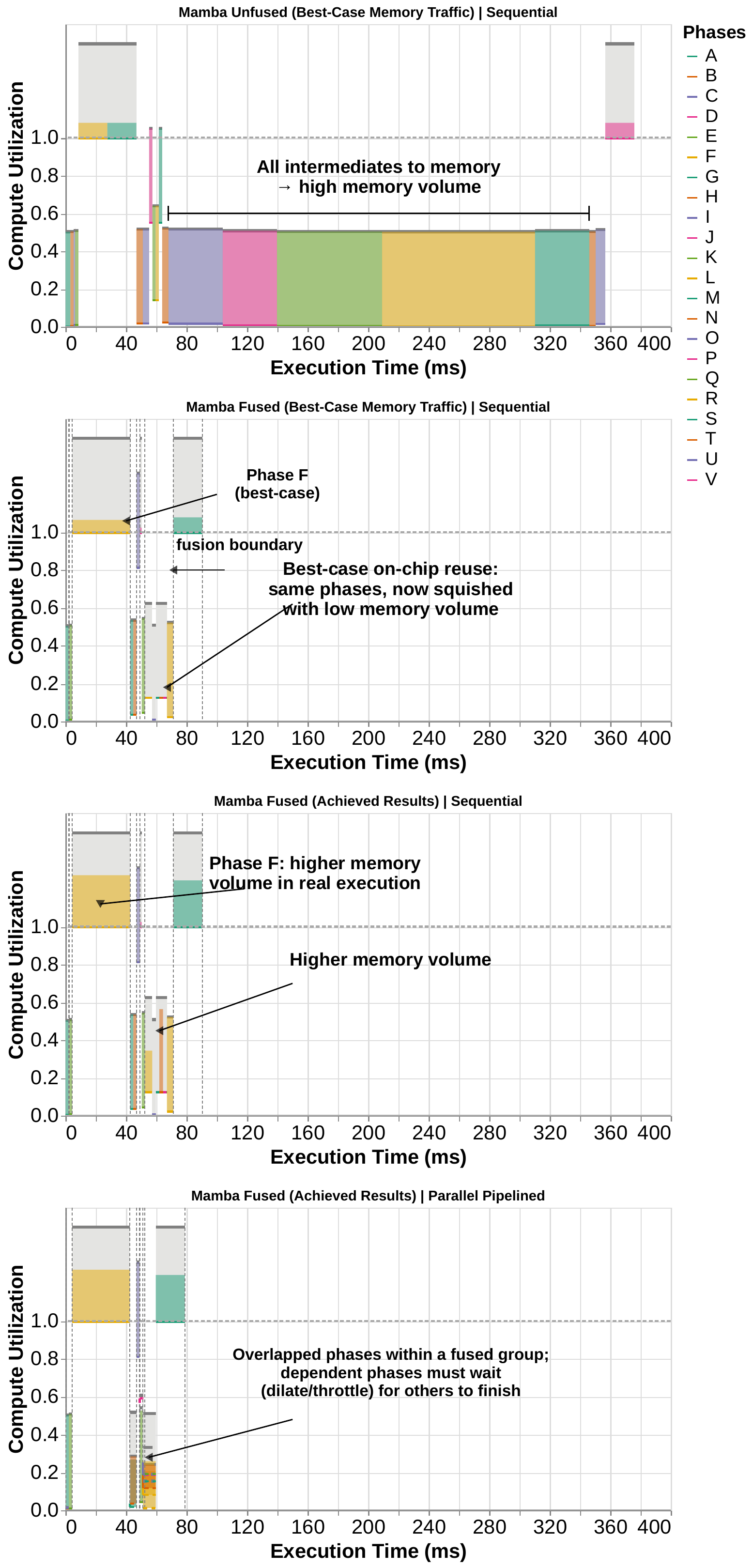}
  \caption{Stacked campaign diagrams for Mamba (top to bottom): unfused,
  fused (analytical/best-case traffic), fused (achieved), and fused with
  parallel pipelining.}
  \label{fig:mamba-campaign-stacked}
\end{figure}

\begin{figure}
  \centering
  \includegraphics[width=\columnwidth]{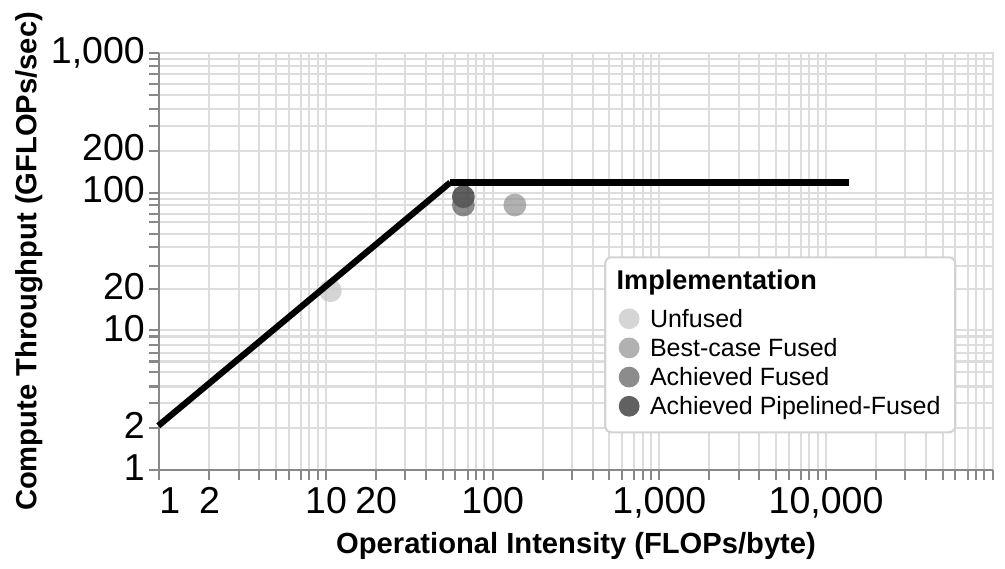}
  \caption{A single roofline with the aggregated points for all four Mamba
  implementation strategies.}
  \label{fig:mamba-rooflines}
\end{figure}

We now turn to a substantially more complex case study: \emph{Mamba-1}.
Mamba is a large language model based on state-space formulations~\cite{Gu:2024:MLT}.
Unlike transformer-based models, which maintain a growing key-value (KV) cache during autoregressive decoding,
Mamba maintains a fixed-size hidden state tensor, akin to a compressed representation of the KV cache.
From an algorithmic perspective, we can precisely describe a single Mamba layer as a sequence of 24 tensor algebra expressions, or \emph{Einsums}~\cite{odemuyiwa:2026:edge, Kjolstad:2017:TAC, Parashar:2019:TSA},
with various data dependencies between them.
These dependencies limit reordering and impose constraints on scheduling and data movement,
making Mamba a challenging workload for both analysis and implementation.

\emph{Baseline analytical model.} We begin by considering a best-case analytical model in which all phases are unfused and executed sequentially. Each phase reads from and writes to main memory.
In this formulation, the dominant bottleneck arises from the state-space model update (middle phases with long latencies in Figure~\ref{fig:mamba-campaign-stacked}, first panel).
The hidden state tensor is large, and each token update requires significant memory traffic relative to the amount of computation performed.
As a result, the unfused Mamba workload is largely memory bandwidth-bound.

\emph{Fusion opportunity.} The campaign diagram for the unfused analytical model suggests that fusing multiple phases into larger execution units may substantially reduce off-chip memory traffic.
Using the analytical model, we evaluate a fused schedule that partitions the 24 phases into 8 fusion groups.
Within each fusion group, intermediate tensors are assumed to be retained on-chip, while off-chip memory accesses are required only at fusion group boundaries.
Figure~\ref{fig:mamba-campaign-stacked} (second panel) visualizes the performance of this idealized fused model.

\emph{Analytical versus simulated behavior.} For a realistic, achieved fused implementation, we use phase results from a Timeloop-based simulation~\cite{Parashar:2019:TSA} on a FuseMax-like accelerator~\cite{Nayak:2024:FML_micro}.
Figure~\ref{fig:mamba-campaign-stacked} (third panel) shows the resulting campaign diagram.
Compared with the idealized fused model (Figure~\ref{fig:mamba-campaign-stacked}, second panel), the simulated results show additional, partially filled memory pipes as total memory traffic is slightly higher.
Perfect data reuse cannot be achieved in practice due to finite on-chip storage and dataflow constraints.
Nevertheless, the overall performance remains near the analytical upper bound due to fusion between phases.

\emph{Pipelining within fusion groups.} Fusion alters the balance between compute throughput and memory bandwidth utilization.
While the unfused workload is almost entirely memory-bound, fusion causes several phases within each fusion group to become compute-bound (Figure~\ref{fig:mamba-campaign-stacked}, third panel).
This shift enables further optimization through parallel pipelining.
By overlapping execution across phases within a fusion group, we can increase overall throughput.
Figure~\ref{fig:mamba-campaign-stacked} (fourth panel) illustrates the impact of this pipelined execution strategy combined with dilation and throttling, which improves end-to-end performance by approximately $4.4\times$ relative to the unfused implementation, and $1.1\times$ relative to the non-pipelined, fused implementation.

\emph{Architectural implications.} Taken together, this sequence of analyses demonstrates how campaign diagrams can guide architectural design.
For Mamba, an architect may prioritize an architecture that
(1) enables fusion across tiles to reduce off-chip memory traffic, and
(2) provides an on-chip network capable of supporting fine-grained pipelining between fused phases.
Campaign diagrams make these trade-offs explicit.
Overall, campaign diagrams scale to realistic, complex workloads and enable comparisons between analytical bounds and achieved implementations.

\section{Conclusion}
This work introduces campaign diagrams, a novel visualization technique that enables phase-level performance reasoning about modern workloads.
We show how fusion changes the memory volume; how pipelining can oversubscribe resources, which campaign diagrams expose along with potential scheduling transformations (e.g., dilation/throttling); and how, counterintuitively, lower operational intensity can still improve end-to-end performance.
Campaign diagrams interface with analytical models, simulations, or profiling data, supporting design-time reasoning before committing to a specific design.

\paragraph{Limitations}
As workloads grow in complexity, diagrams can become visually dense and harder to interpret. Our tool supports zooming, but presenting such views statically (e.g., on paper) remains a challenge.
We currently support any pair of rate-limited resources; extending beyond two is future work.

\section*{Acknowledgements}
We thank Serban Porumbescu, Chris Fletcher, Nandeeka Nayak, Yan Zhu, and Michael Gilbert for their feedback and technical discussions on campaign diagrams and their applications.
This work was supported in part by an NVIDIA Graduate Fellowship and NSF Grant 2403389.

\bibliographystyle{template/IEEEtranS}
\bibliography{refs, related, clean_mamba, dissertation}

\end{document}